\documentclass[manuscript]{aastex}
\usepackage{graphicx}
\usepackage{natbib}

\shorttitle{Properties of quiet Sun and Active Regions}
\shortauthors{Criscuoli}

\begin{document}

   \title {Comparison of physical properties of quiet and active regions through the analysis of MHD simulations of the solar photosphere
    }
   \author{S.~Criscuoli\altaffilmark{1}}

\altaffiltext{1}{ National Solar Observatory, Sacramento Peak, P.O. Box 62, Sunpsot, NM 88349, USA}

   \date{}

  \begin{abstract}
Recent observations have shown that the photometric and dynamic properties of granulation and of small-scale magnetic features depend on the amount of magnetic flux of the region they are embedded in. We analyze results from numerical Hydro and Magneto Hydrodynamic simulations characterized by different amount of average magnetic flux and find qualitatively the same differences as those reported from observations. We show that these different physical properties result from the inhibition of convection induced by the presence of the magnetic field, which changes the temperature stratification of both quiet and magnetic regions. Our results are relevant for solar irradiance variations studies, as such differences are still not properly taken into account in irradiance reconstruction models. 
   
\end{abstract} 


\maketitle

\section{Introduction}
\label{sec:intro}
Solar irradiance variations measured by radiometers can be reproduced by ad-hoc modeling of physical properties of quiet Sun and magnetic features, and the use of sophisticated radiative transfer codes \citep[][and references there in]{ermolli2013}. The accuracy of such reproductions depends on the wavelengths at which irradiance is measured and on the temporal scales considered. For instance, models can reproduce more than 90$\%$ of the Total Solar Irradiance (the irradiance integrated over the whole electromagnetic spectrum) variations measured from days to the magnetic solar cycle, while the agreement is less good when considering longer temporal scales \citep{frohlich2011} or the Spectral irradiance \citep[that is the irradiance integrated over particular wavelength regions, e.g.][]{harder2009}.
Such discrepancy suggests that the employed one-dimensional and static atmosphere models describing the properties of features observed on the Sun need to be improved.

For instance, the role of the quiet Sun is still unclear. For irradiance reconstructions, its properties are considered not to vary with the magnetic activity, but recent measurements \citep{criscuoli2013} as well as MHD numerical simulations \citep{fabbian2010, fabbian2012} have shown that a small amount of unresolved magnetic flux (average less then 100~G) can modify the shape of some magnetically sensitive photospheric Iron lines, thus contributing to modulation of the spectral irradiance  signal.  Moreover, such models still not properly take into account the dependency of the properties of granulation on the average magnetic flux of the regions in which it is located.  For instance, \citet{kobel2012}, in agreement with results presented previously by \citet{narayanscharmer2010} and \citet{ishikawa2007}, found that the rms of Line of Sight (LOS, hereafter) velocities is reduced with the increase of the magnetic flux not only in moderate flux areas, but also in relatively quiet areas (flux lower than 100~G). \citet{morinaga2008}, suggested that convective motions in magnetic regions are mostly inhibited by the presence of small-scale magnetic features, whose presence hinders horizontal motions and therefore also vertical ones \citep[see also][]{keller1991}. \citet{baudin1997} showed that even horizontal convective motions of granulation embedded in a facular region are reduced with respect to motions found in quiet regions. The presence of magnetic field also determines the height at which convection overshoots. \citet{kostik2012}  found for instance that within magnetic structures convection overshoots at higher layers. 
As convection strongly contributes to the transport of energy at the base of the photosphere, these results suggest that such dependence should be taken into account for proper solar irradiance reconstruction. Indeed, a clear anti correlation between the intensity contrast of granulation and the magnetic flux has been reported by several authors \citep[e.g.][]{kobel2012, kostik2012,baudin1997, montagne1996, title1992}.

Observations have also shown that photospheric bright points observed in quiet Sun regions show higher photometric contrast \citep[e.g.][]{lawrence1993,ishikawa2007,kobel2011,romano2012,feng2013} and larger LOS velocities \citep[e.g.][]{ishikawa2007,narayanscharmer2010,romano2012} then those observed in ARs. Irradiance reconstruction techniques, especially those based on analysis of magnetograms \citep[][,and references therein]{ball2011}, still do not take such differences into account.

At the best of our knowledge, few works have employed results from  Magneto Hydro Dynamic (MHD) simulations to investigate  the differences of physical properties of plasma embedded in varying magnetic flux typical of quiet and facular regions.  
\citet{cattaneo2003} showed that convective motions are more and more inhibited as the average magnetic flux in their snapshots is increased. \citet{vogler2005a} showed that the average bolometric intensity contrast varies non monotonically with the increase of the average flux in the simulations.

In this paper we employ results from MHD simulations to investigate and compare the properties of granulation and of magnetic features embedded in regions characterized by different average values of magnetic flux. We show that MHD simulations reproduce, at least qualitatively, the differences found in observations, and put forward a physical explanation. 

The article is organized as follows: in Sec. 2 we describe the simulations and the analysis performed; in Sec. 3  we present our results, which are compared to observations in Sec. 4; in Sec. 5 we show that the dependence of properties of plasma on the amount of magnetic flux of the surrounding regions is caused by the changes of the average temperature stratifications induced by the suppression of convection by the magnetic field; in Sec. 6 we summarize our findings and draw our conclusions.


\section{Simulations and analysis}
The study was performed analyzing several 3D MHD snapshots obtained with the Copenhagen-Stagger code \citep{nordlund1995}. In particular, we considered a set of 10 HydroDynamic (HD, hereafter) snapshots, and three sets of 10 snapshots characterized by average magnetic flux values of 50, 100 and 200~G. The original snapshots simulated a portion of 6$\times$6 Mm$^2$ of the solar photosphere, horizontally sampled at 24~ km/pixel; the vertical resolution was not constant but equal to about 15~km/pixel at the base of the photosphere; each snapshot had original dimensions of 252x252 in the horizontal direction and 126 pixels in the vertical one. For our analysis we considered a portion of the original snapshots  along the vertical direction,  which included 500 km above and below (for a total of 1~Mm) the height that corresponded to average optical depth unity at 500 nm in the HD snapshots. The snapshots are described in detail in \citet{fabbian2010,fabbian2012}. We then employed the RH code \citep{uitenbroek2002,uitenbroek2003} to 
synthesize the emergent continuum intensity radiation at 608 nm and to compute the optical depth at 500 nm ($\tau_{500}$, hereafter) in all the snapshots.
Physically relevant quantities, as plasma velocities, magnetic field intensities and temperature were computed at different optical depth surfaces by interpolation on the vertical grid. Namely we considered surfaces at $\tau_{500}$ = 1, 0.3 and 0.1, which correspond to average geometric heights of 0, -70 and -150 km respectively, in the HD snapshots (negative values correspond to heights above the average optical depth unity surface). The average height at $\tau_{500}$ = 1 is on average shifted downward by approximately 2, 8 and 20 km in the 50, 100 and 200~G simulations, respectively. Note that the horizontal velocity was computed as square root of the sum of the squares of the velocities along the two horizontal directions divided by two.

To clearly distinguish between the average magnetic flux of the MHD snapshots from that of magnetic features observed within them, we will address the former as \textit{environmental magnetic flux} in the following.   

\begin{figure*}
\epsscale{.3}
\includegraphics[trim = 0mm 15mm 55mm 5mm, clip, width=4.6cm, height=4.0cm]{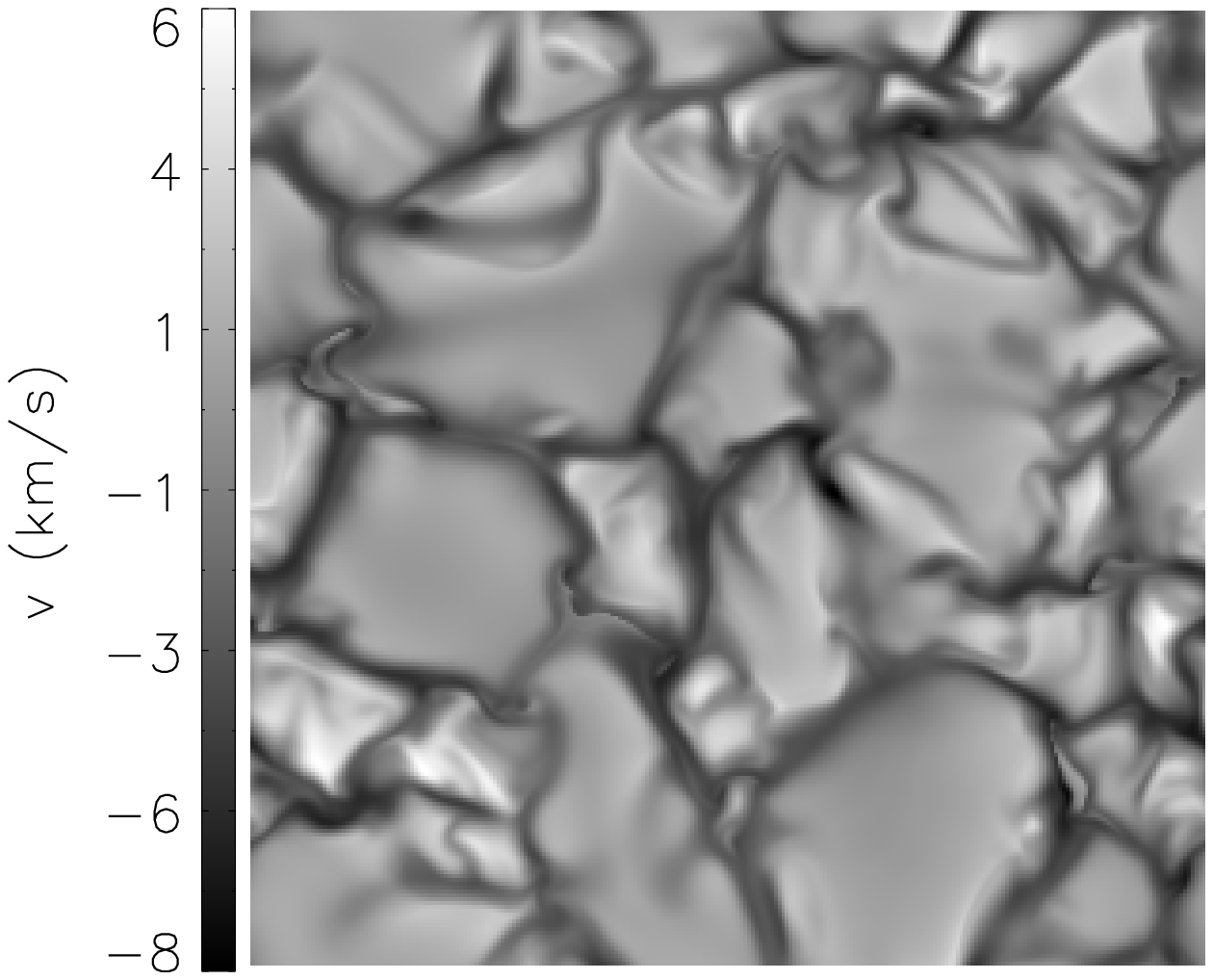}
\includegraphics[trim = 0mm 1mm 78mm 5mm, clip, width=3.6cm]{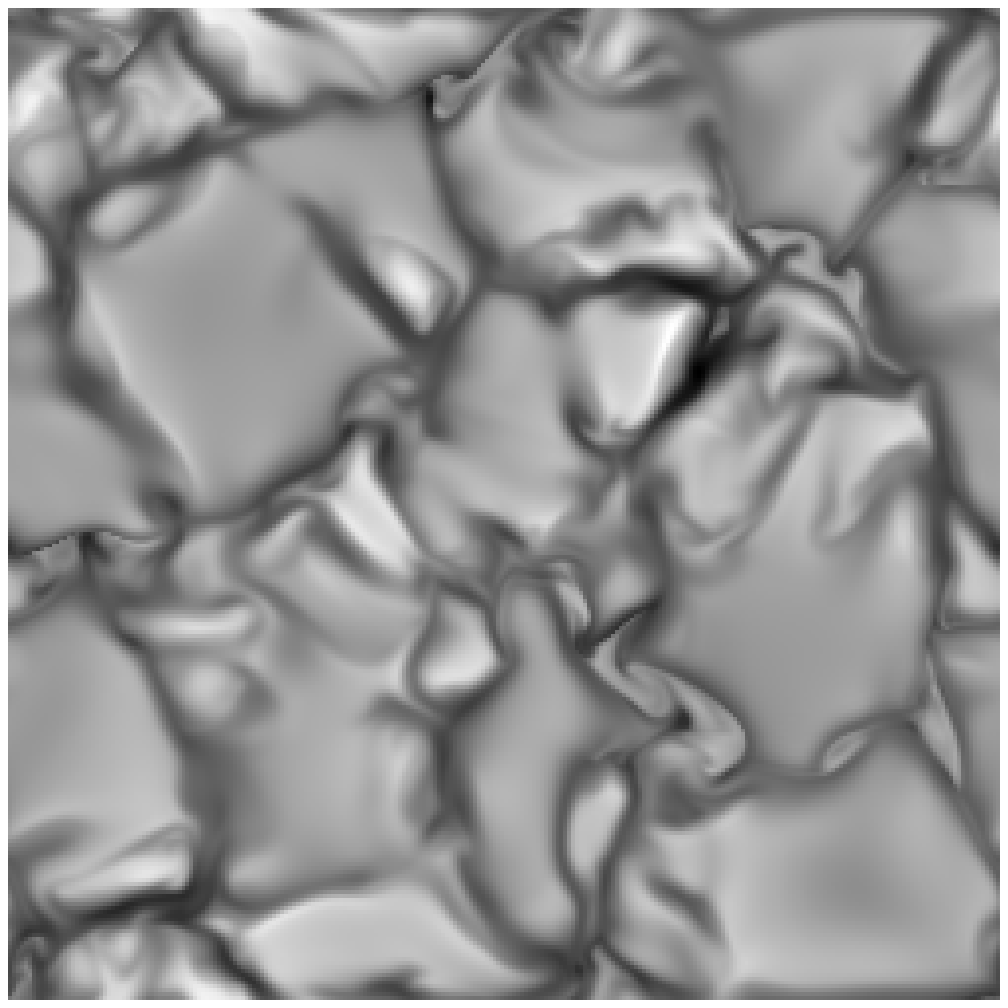}
\includegraphics[trim = 0mm 1mm 78mm 5mm, clip, width=3.6cm]{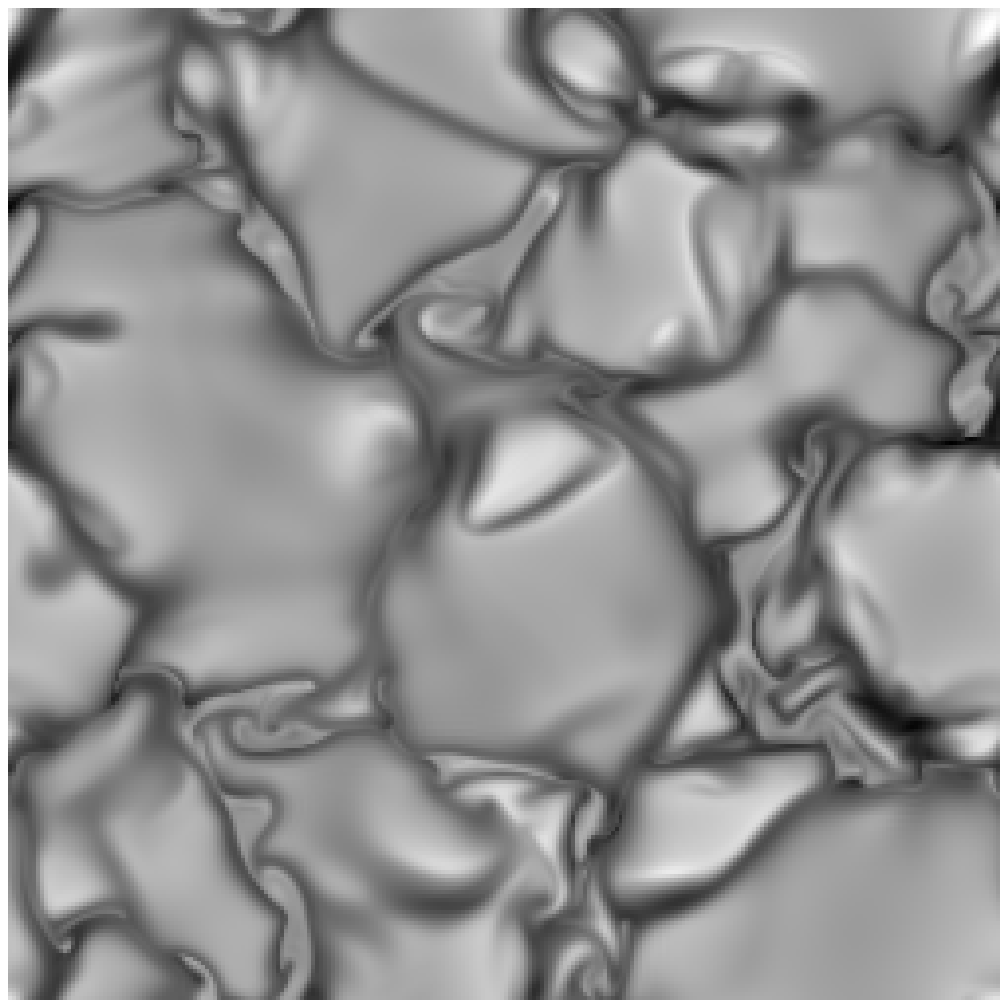}
\includegraphics[trim = 0mm 1mm 78mm 5mm, clip, width=3.6cm]{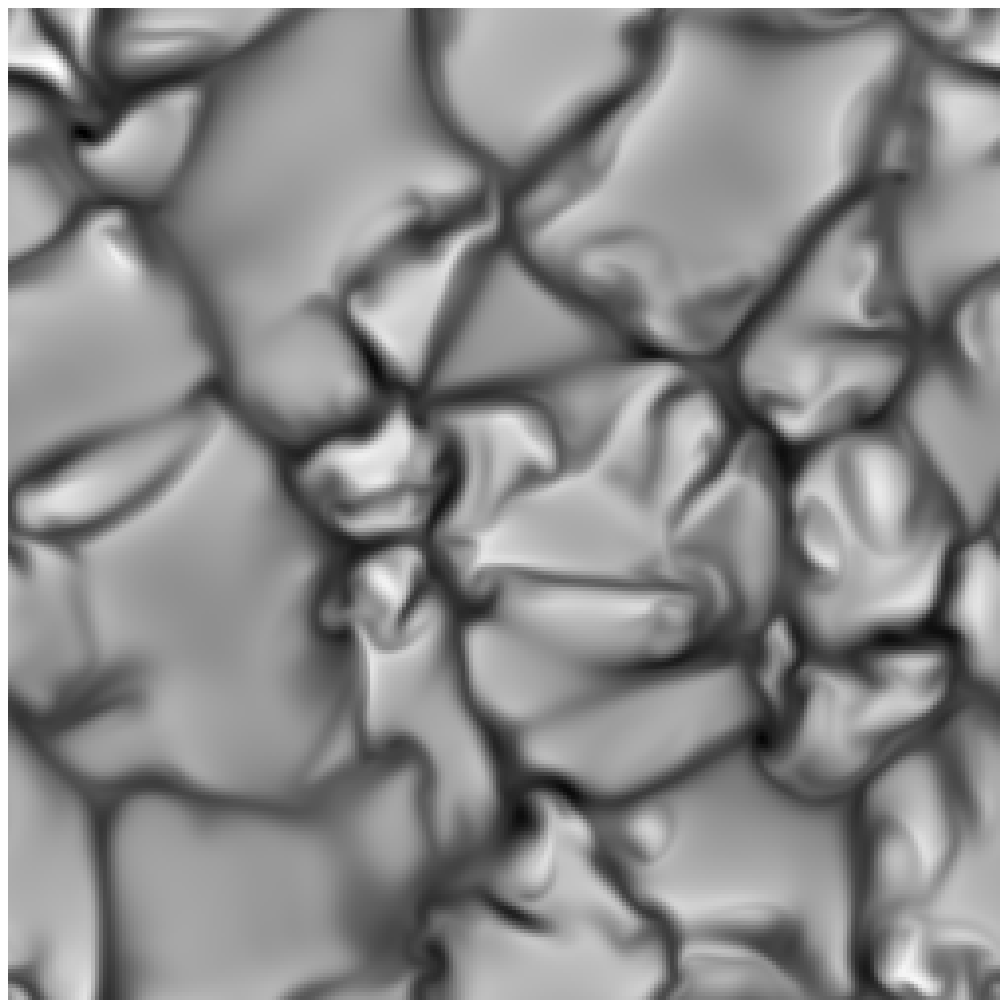}\\

\includegraphics[trim = 0mm 15mm 55mm 5mm, clip, width=4.6cm, height=4.cm]{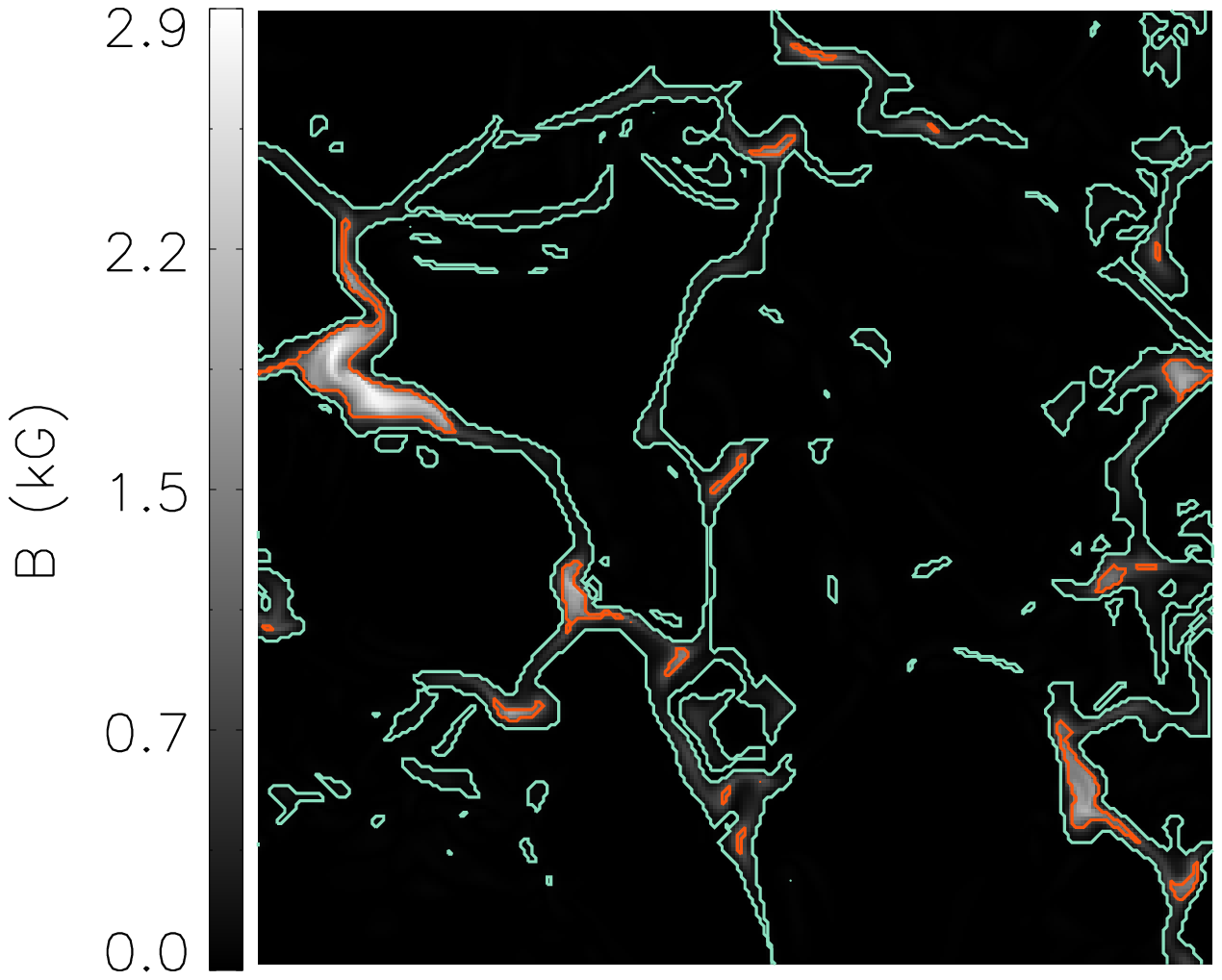}
\includegraphics[trim = 0mm 1mm 78mm 5mm, clip, width=3.6cm]{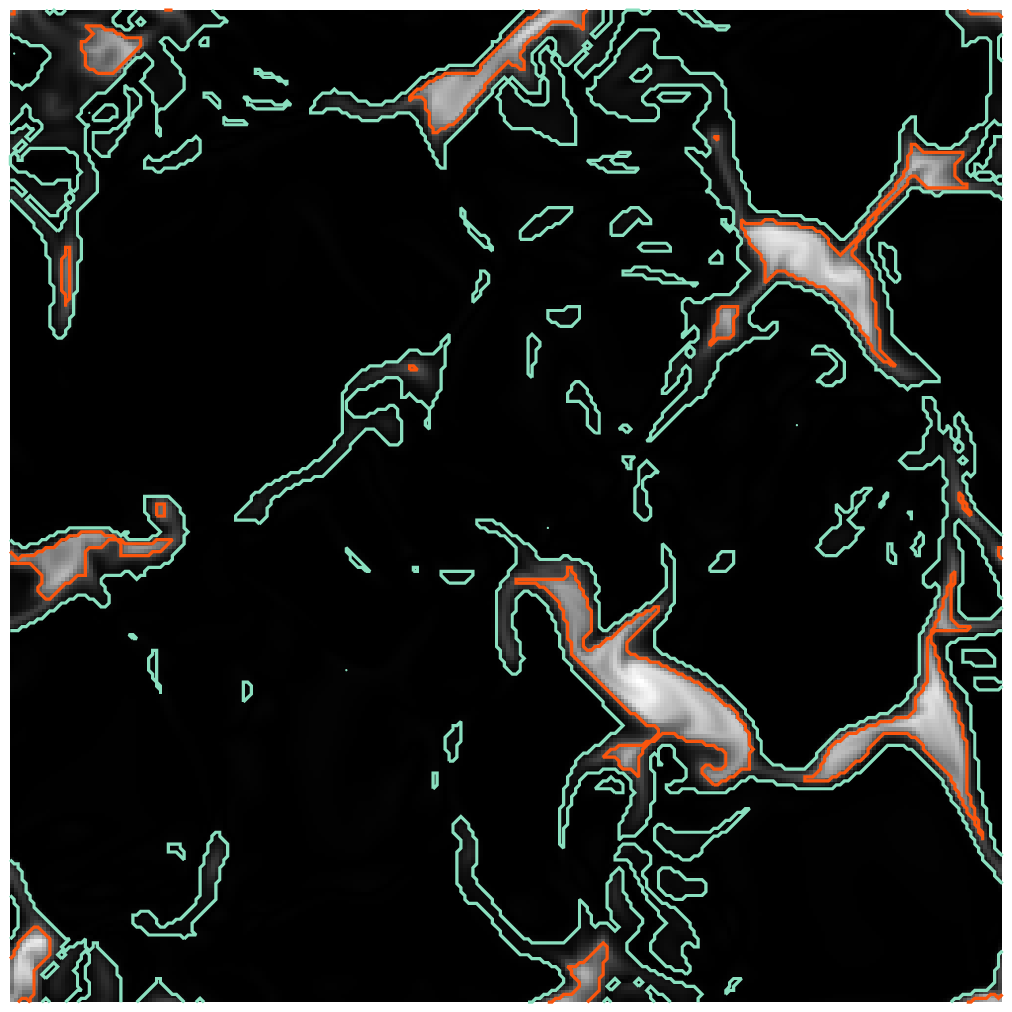}
\includegraphics[trim = 0mm 1mm 78mm 5mm, clip, width=3.6cm]{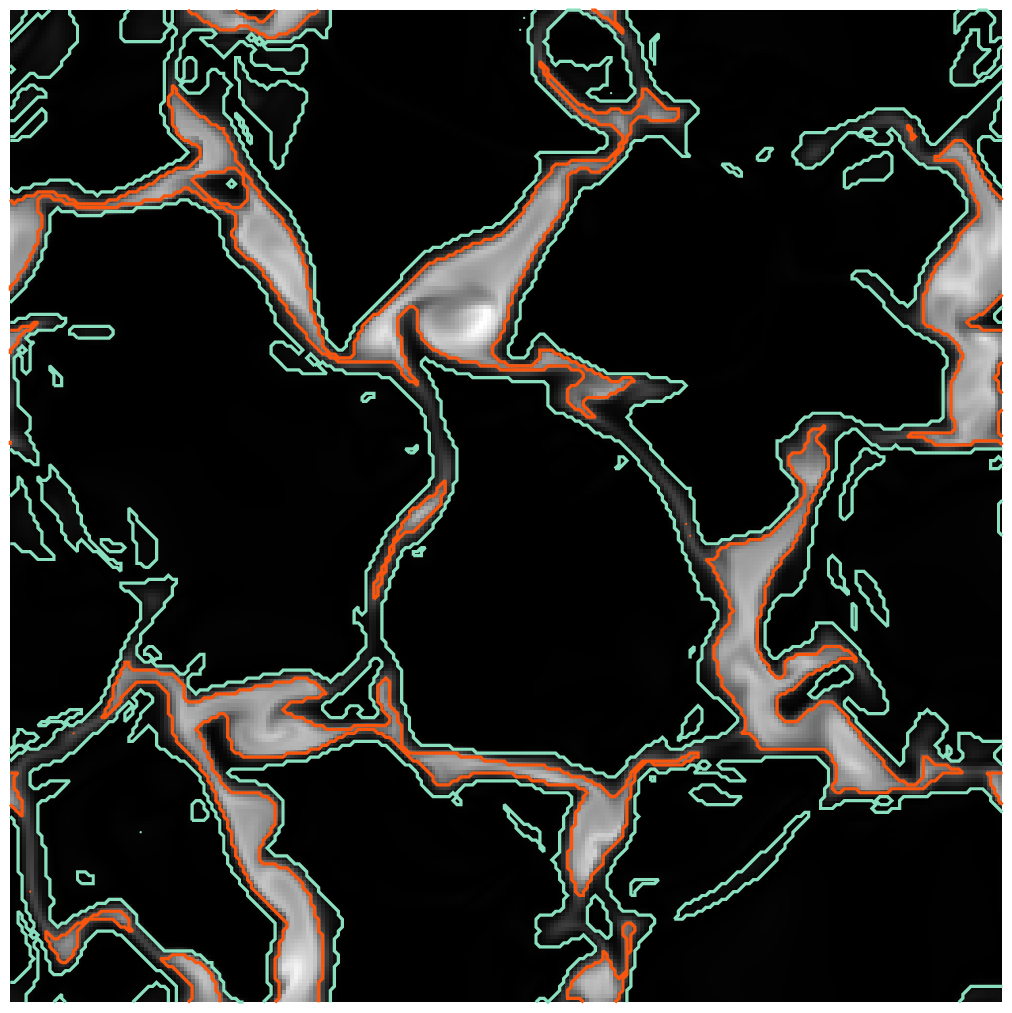}\\

\caption{Top: Examples of vertical velocity fields at $\tau_{500}$=1 from the (M)HD snapshots;  negative velocities correspond to downflows. Bottom: corresponding absolute intensity of vertical component of magnetic field; red contours mark borders of regions where absolute vertical field intensity is larger than 800~G; blue lines mark borders of regions where vertical field intensity is smaller than 50~G. From left to right, examples of 50~G, 100~G, 200~G and HD snapshots. 
\label{panels_examples}  }  
\end{figure*}

Figure \ref{panels_examples} shows examples of the vertical component of the velocity field (top) and of the corresponding absolute intensity of the vertical component of the magnetic field (bottom) at $\tau_{500}$=1. It is interesting to notice that, in agreement with \citet{cattaneo2003}, with the increase of the environmental magnetic flux there is an increase of the filling factor of large magnetic structures  which form at the vertexes of granules. These features \citep[addressed by some authors as micropores,][]{vogler2005a} have typical field intensity of the order of 1-2 kG and velocities within them are reduced with respect to their surroundings. Figure \ref{panels_examples} also shows that these large features can be substructured, as fluctuations of magnetic field intensity, velocities and temperature (not shown) are observed within them.  We also note that the 50~G simulations present a larger number of filamentary magnetic structures located within downflow regions at the edges of granules, which have typical field strengths of some hundred G. We also notice the presence of strong down-flows (up to approximately -7 km/s) at the edges of magnetic structures, which are driven by radiative losses due to the reduction of opacity inside magnetic features \citep[e.g.][]{steiner1998,lekasteiner2001,vogler2005a}.
\label{sec:simu}

 To distinguish between properties of magnetic structures and granulation we defined as  \textit{non magnetic pixels} and as \textit{magnetic pixels} those regions where the absolute intensity of the vertical component of the magnetic field at $\tau_{500}$ = 1 is smaller than 50~G, and larger than 800~G, respectively. The contours of these regions are marked in blue and orange, respectively, over the magnetic field strength images in Fig.\ref{panels_examples} (bottom panels). We notice that in this way we automatically  selected the features characterized by kG field strength, located at the borders of granules, described above.
  
To investigate the differences of physical properties of convective plasma observed in different environments we analyzed  the shapes of the distributions, in particular the standard deviation,  of the vertical and horizontal velocities and of the temperature at different optical depths.\textbf{ We analyzed these quantities because the fluctuations of the vertical component of the velocity and of the temperature are proportional to the convective flux \citep[e.g.][]{stix, brandenburg2005}, therefore these quantities have been often employed as indicators of the vigor of convection \citep[][and references therein]{nesis2006}. } Finally, we investigated the relation between the magnetic field intensity and the continuum contrast at 608 nm. We chose this wavelength because it is in a region of the solar spectrum far from absorption lines \citep[see][]{uitenbroek2013} and is close to 630 nm continuum range investigated by \citet{kobel2012} and \citet{kobel2011}.


\section{Results}
In order to favor a qualitative comparison with observations, in the following we present results obtained at given optical depths.  For completeness, we also investigated variations of physical quantities with geometric height (not shown) and found results qualitatively in agreement with those obtained at equal optical depths. The only exception is the depth dependence of  temperature fluctuations within\textit{ magnetic pixels}. The reasons of this discrepancy are discussed in Sec.~\ref{sec:temper}.

\subsection{Convective motions}
\label{sec:conv}
\begin{figure*} 
\includegraphics[width=5cm]{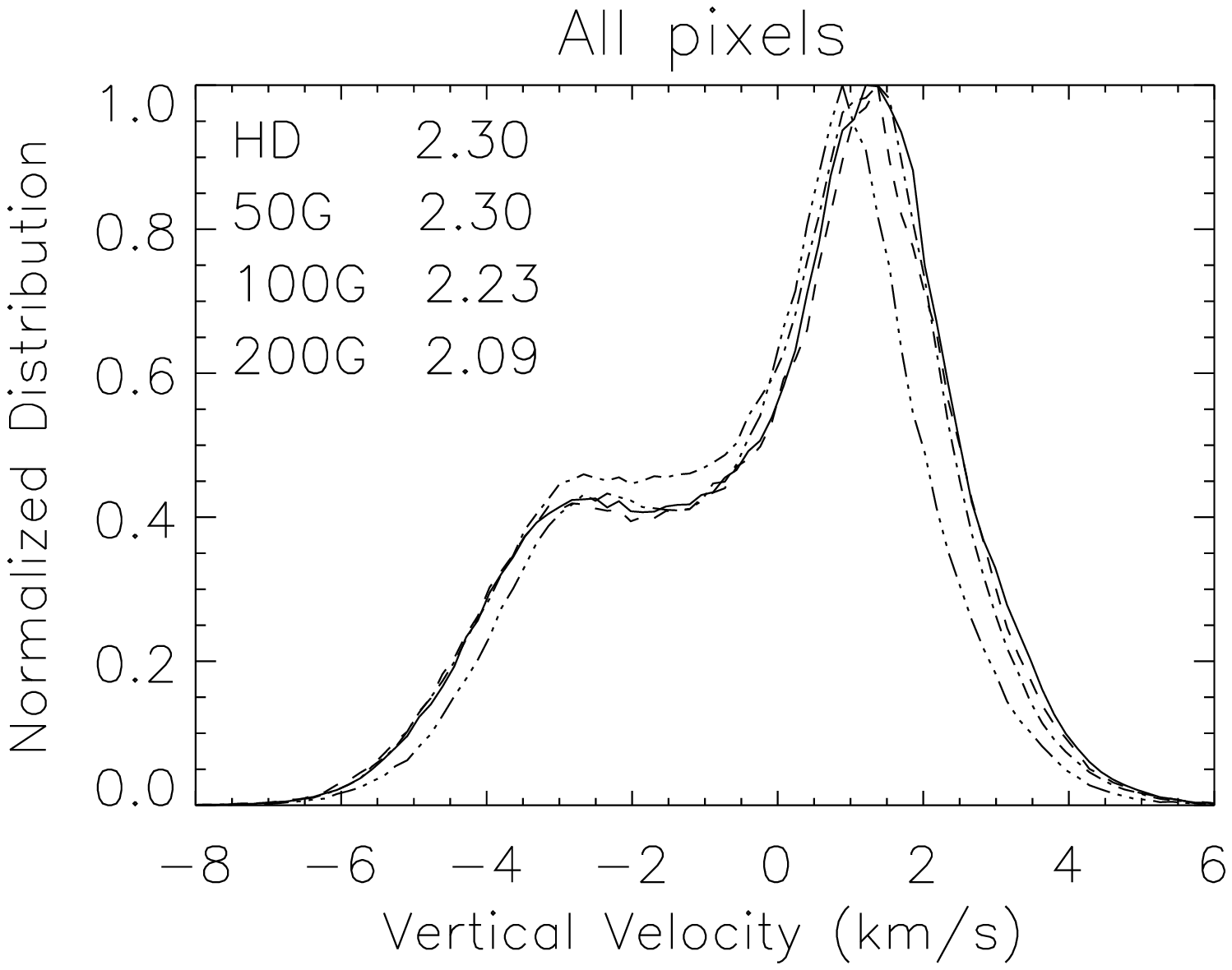}
\includegraphics[width=5cm]{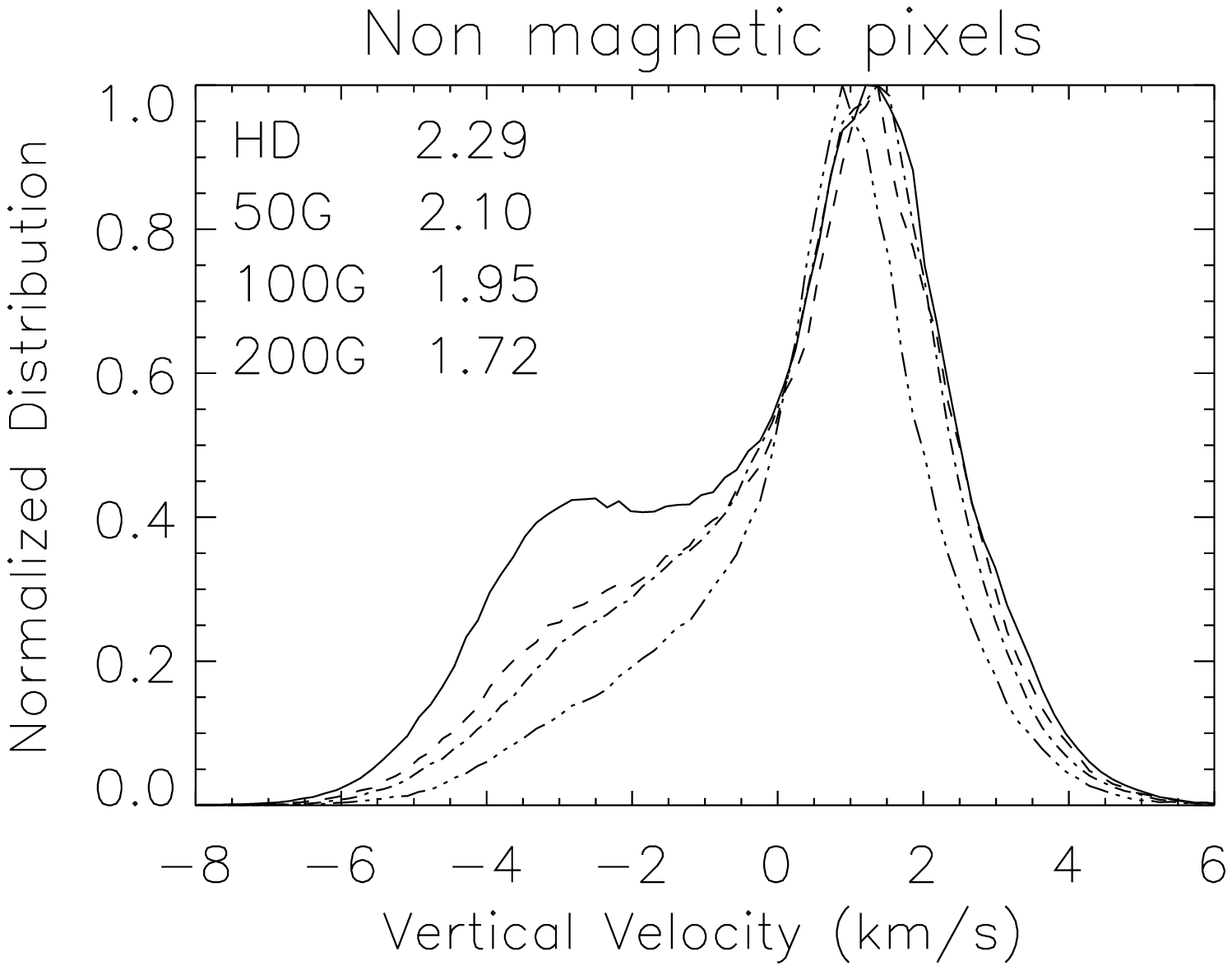}
\includegraphics[width=5cm]{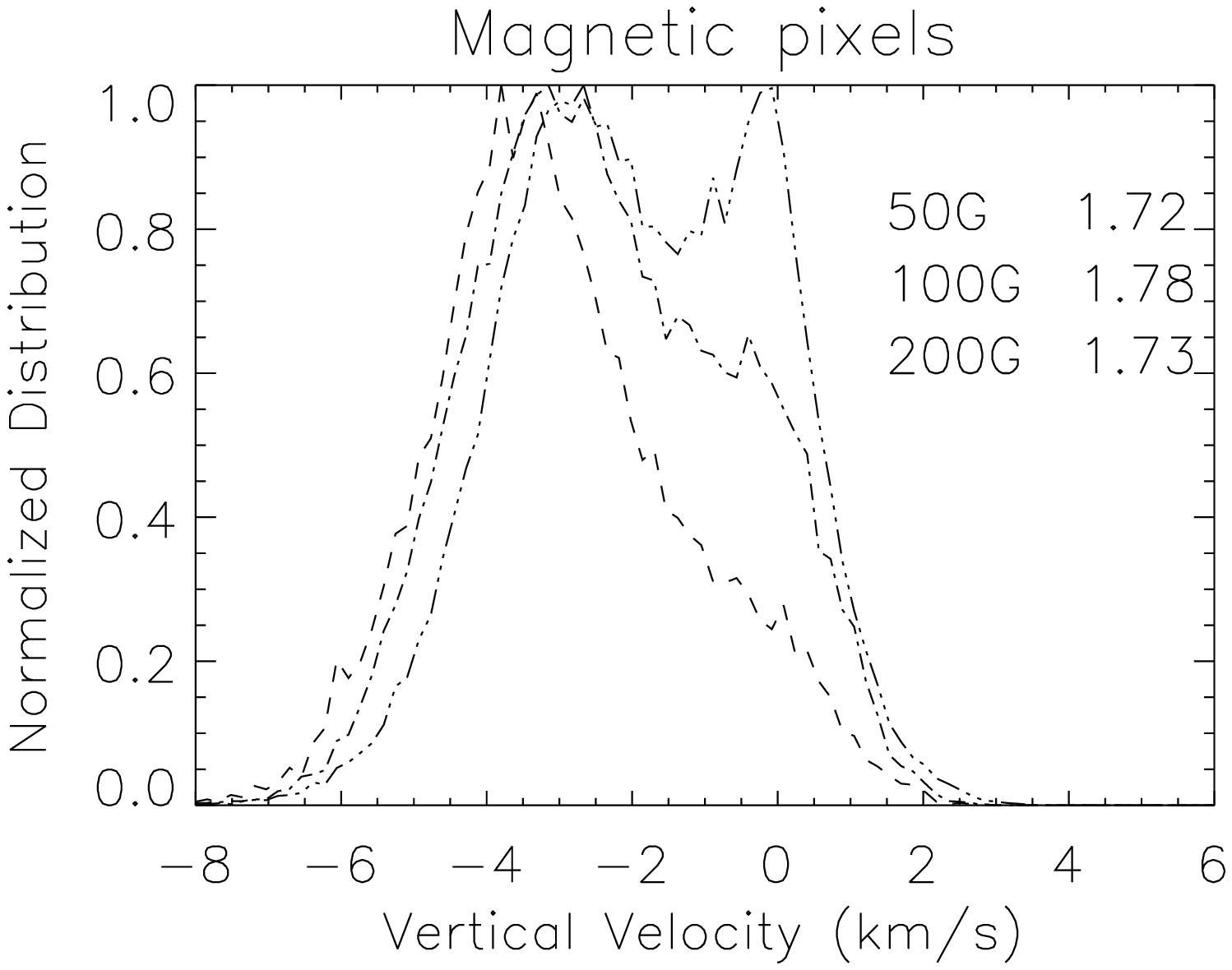}\\

\includegraphics[width=5cm]{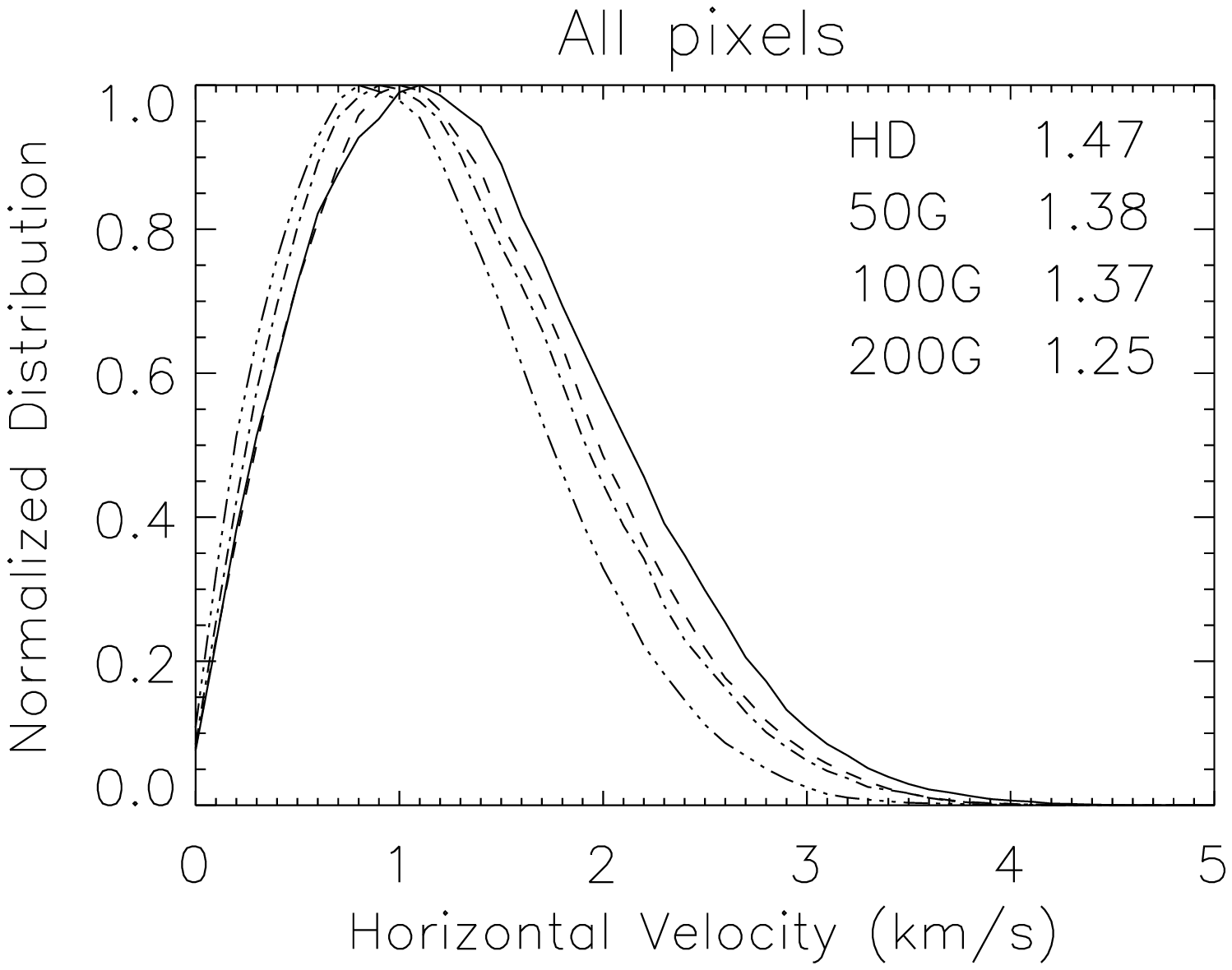}
\includegraphics[width=5cm]{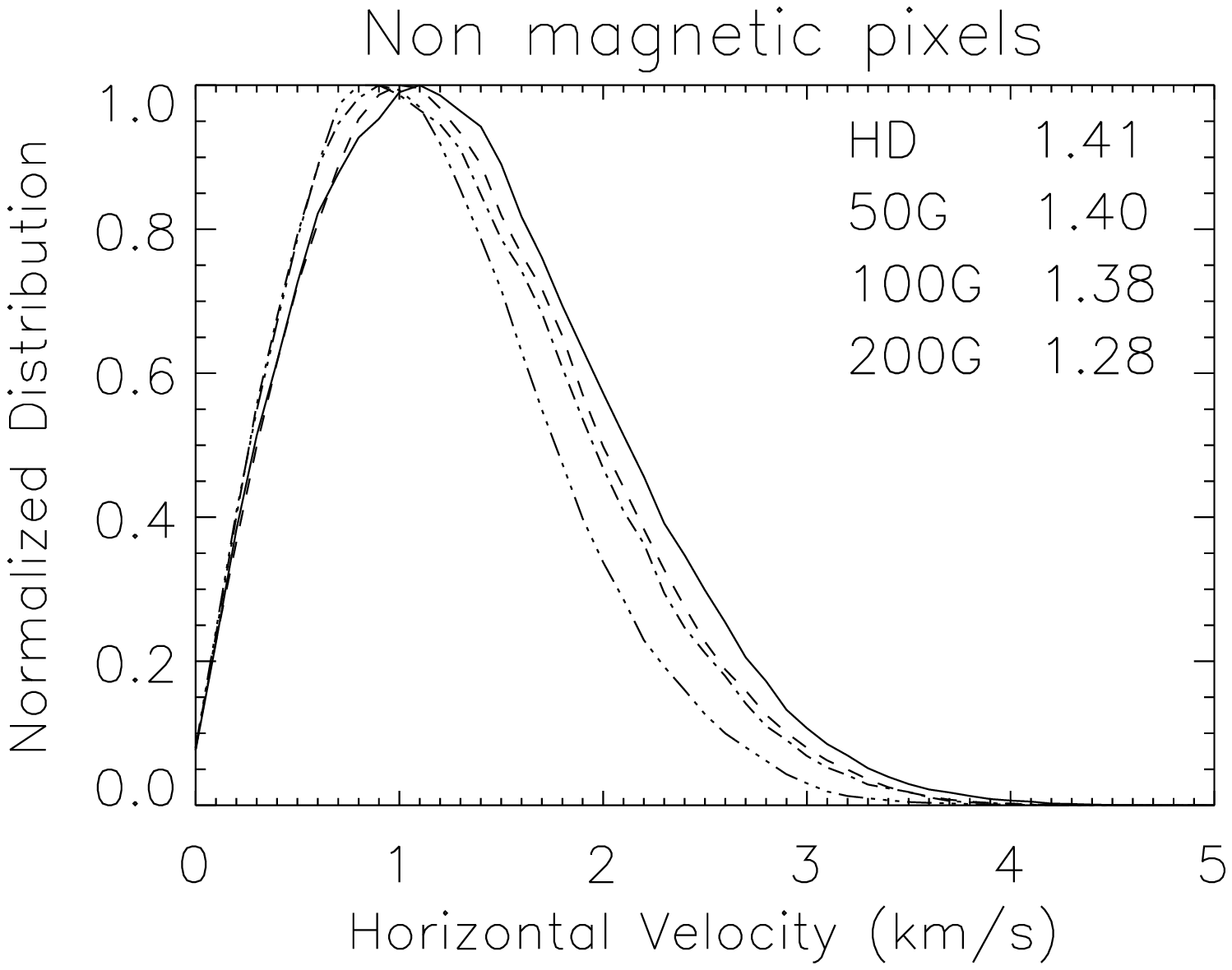}
\includegraphics[width=5cm]{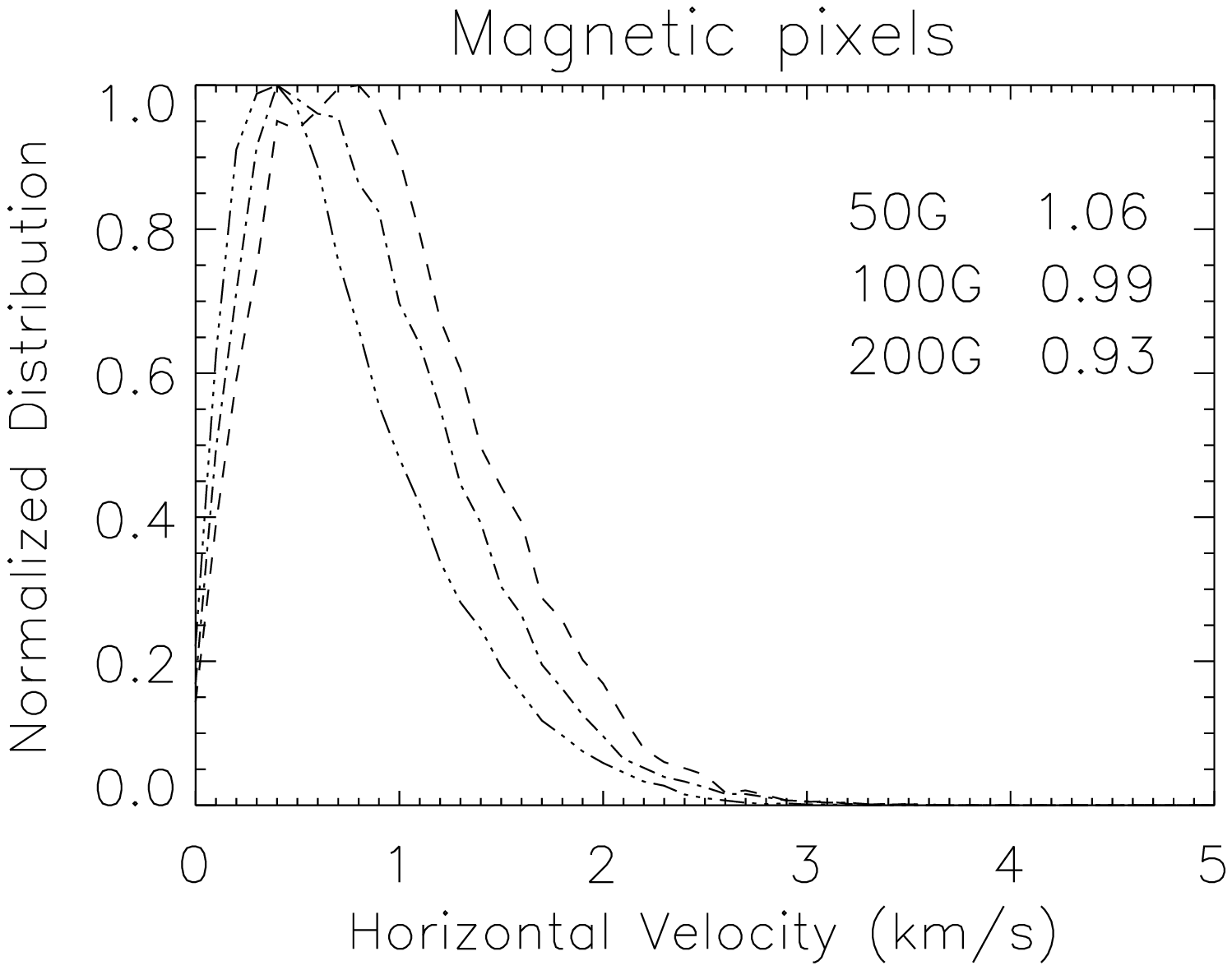}\\

\caption{Velocity distributions at  $\tau_{500}$ = 1 \textbf{and the corresponding standard deviation values}. Top: Vertical velocities. Bottom: Horizontal velocities. Positive values correspond to upflows, negative to downflows. Continuous: HD; dashed: 50~G; dot-dashed: 100~G; dot-dot-dashed: 200~G. 
\label{V_vert_tau1}  }

\end{figure*}

Figure \ref{V_vert_tau1} shows the distributions of vertical (top) and horizontal (bottom) velocities at $\tau_{500}$ = 1 obtained from simulations for different environmental magnetic flux values, together with the corresponding standard deviation values. 
 
We found that, in general, the increase of the environmental magnetic flux causes the distributions to narrow and their peaks to shift toward lower (absolute) velocity values. 
The distributions of vertical velocities of \textit{all pixels} in the snapshots (top left), is less affected by the increase of the environmental magnetic flux.  In this case we clearly distinguish double peaked distributions, which correspond to granular  and intergranular regions. The peak of upflow velocities (granules) occurs at approximately  1.5 km/s in all cases but for the 200 G case, for which the peak occurs at approximately 0.9 km/s. Similarly, the peak of downflow velocities (intergranular lanes) occurs at approximately -3 km/s in all snapshots but for the 200 G one, for which the peak occurs at approximately -2.5 km/s.  Significant differences in the distributions are instead found when comparing velocities of \textit{non magnetic} pixels (top middle), as the percentage of non magnetic pixels that have associated downflows is strongly reduced even in the 50~G case. This is due to the fact that, since magnetic flux is mostly located in intergranular lanes, in MHD simulations these features are more likely to be classified as \textit{magnetic pixels}.  Changes of the shape of the distributions of vertical velocities induced by the increase of the environmental magnetic flux are quite evident in the case of \textit{magnetic pixels}. For these pixels, the distributions show a double peak, one between -4 km/s and -3 km/s (depending on the amount of the environmental magnetic flux) and one at about 0 km/s. These peaks correspond to the large downflow structures found at the edges of magnetic features, and to the pixels almost at rest within them (cfr. Sec.~\ref{sec:simu}), respectively.  With the increase of the environmental magnetic flux, therefore, an increase occurs of the relative number of \textit{magnetic pixels} where plasma velocities are mostly suppressed. 

 A closer analysis of the distributions shows interestingly that vertical velocities of \textit{non magnetic pixels} are slightly skewed toward higher downward velocities, and that this asymmetry slightly increases with the increase of the environmental magnetic flux. Conversely, velocities of magnetic pixels are skewed toward higher upflows velocity values, with the skewness value decreasing with the increase of the environmental flux. 

Bottom panels in Fig.~\ref{V_vert_tau1} show that the shapes of the distributions of the horizontal velocities are asymmetric and show long tails toward higher velocity values. Even in this case the peaks of the distributions shift toward lower velocity values and the widths decrease with the increase of the environmental magnetic flux.

We then investigated the standard deviations of velocity distributions  at different equal optical depths surfaces.

\begin{figure} 
\includegraphics[width=5cm]{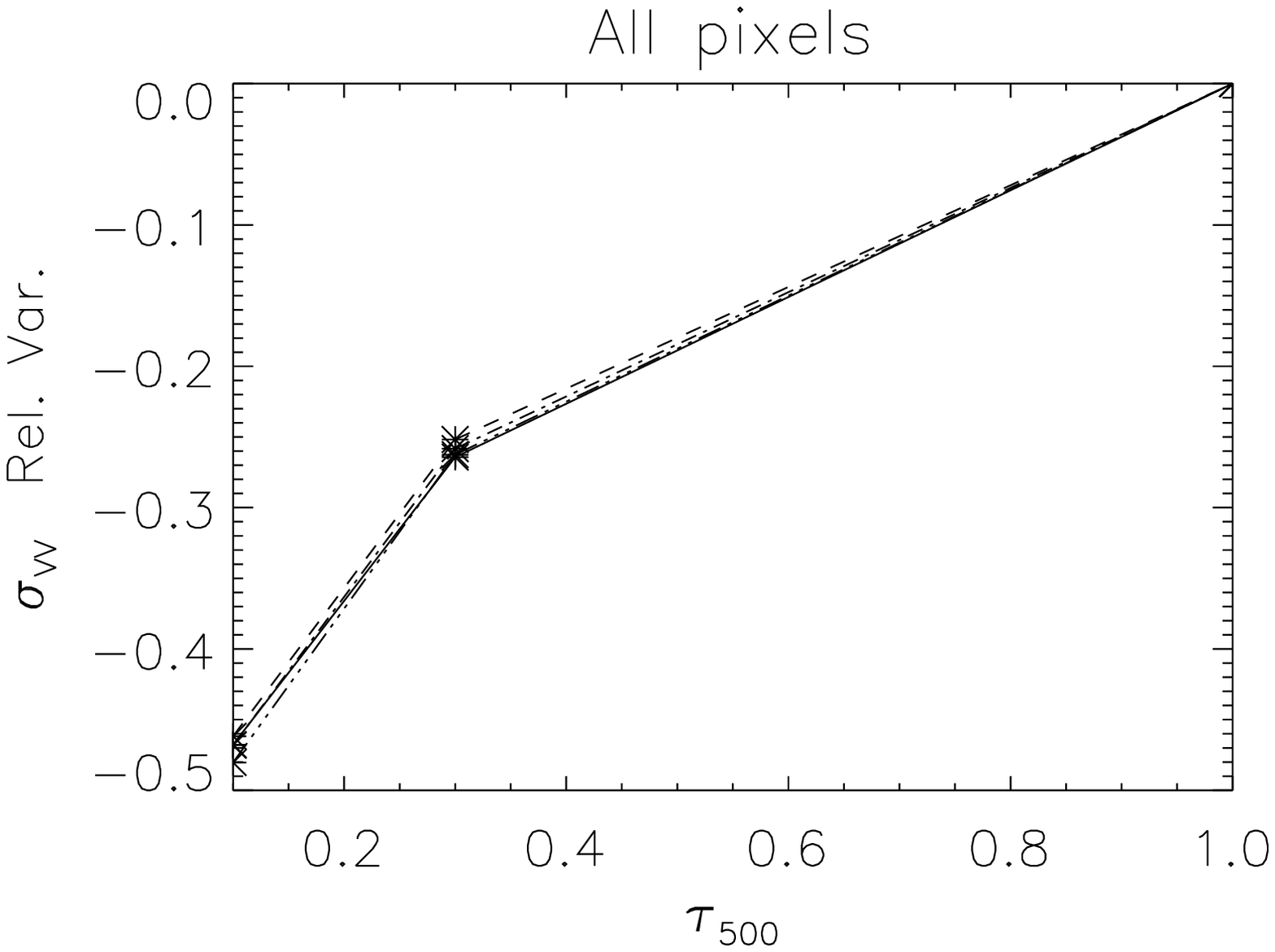}
\includegraphics[width=5cm]{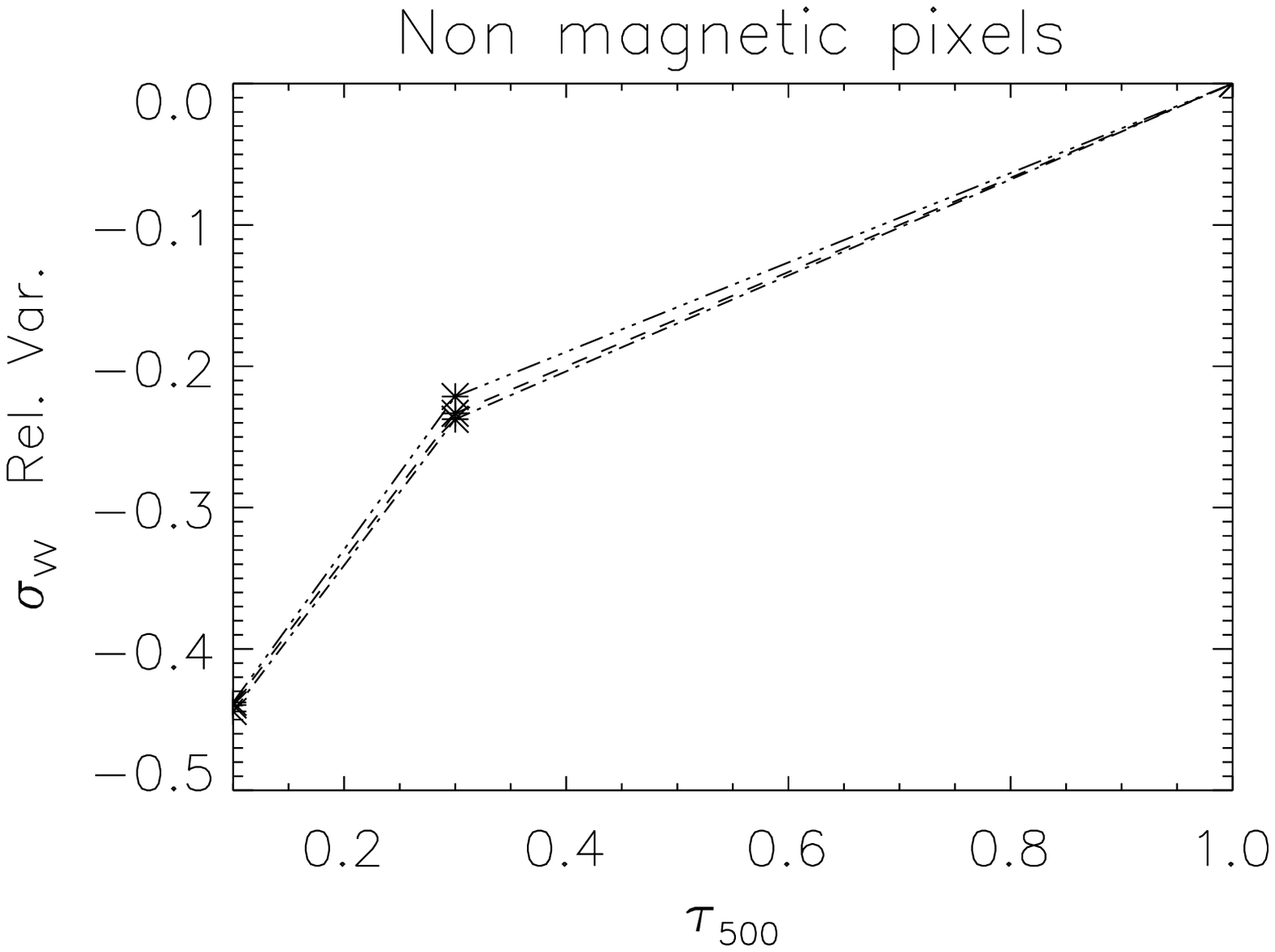}
\includegraphics[width=5cm]{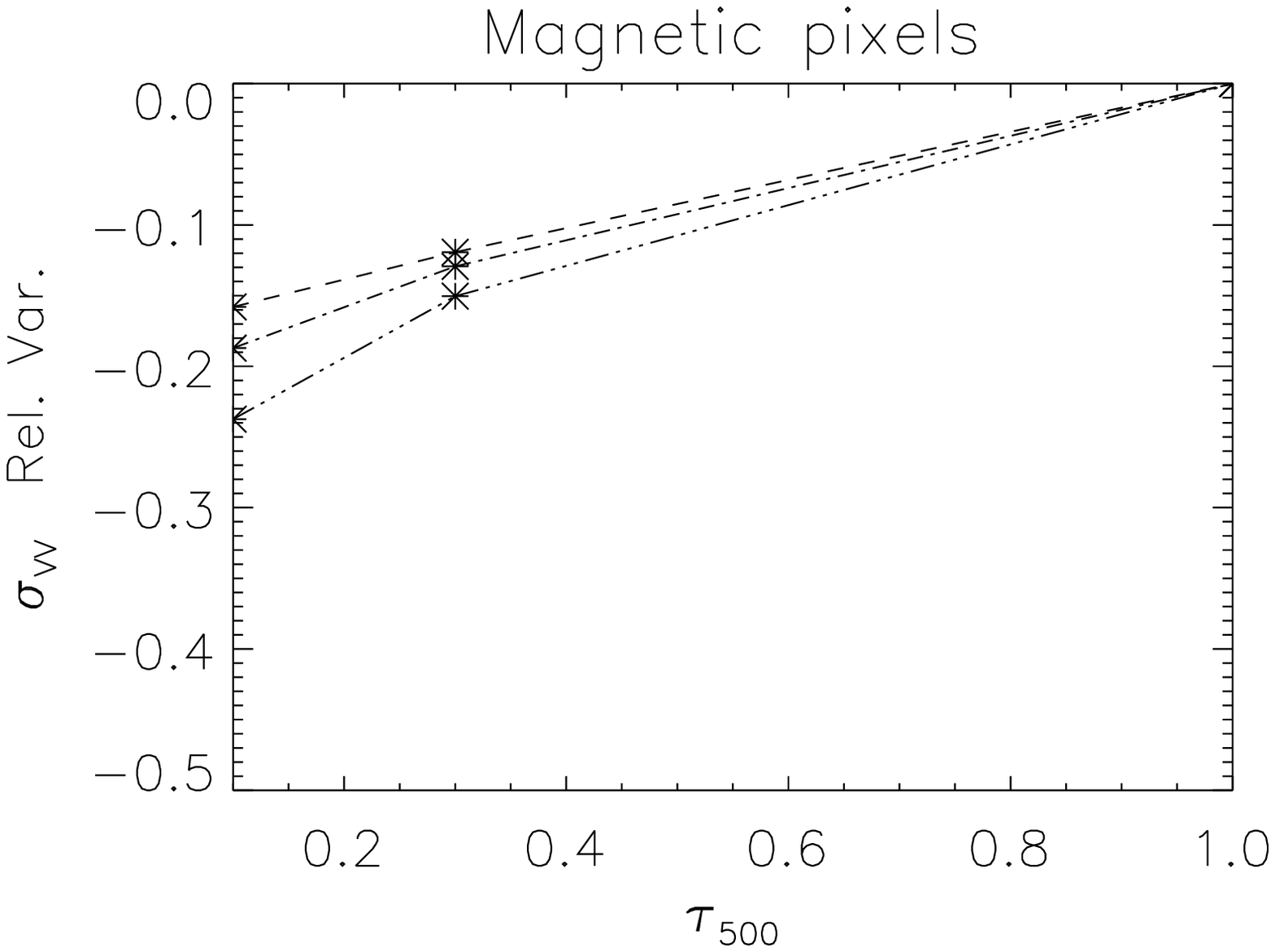}\\
\includegraphics[width=5cm]{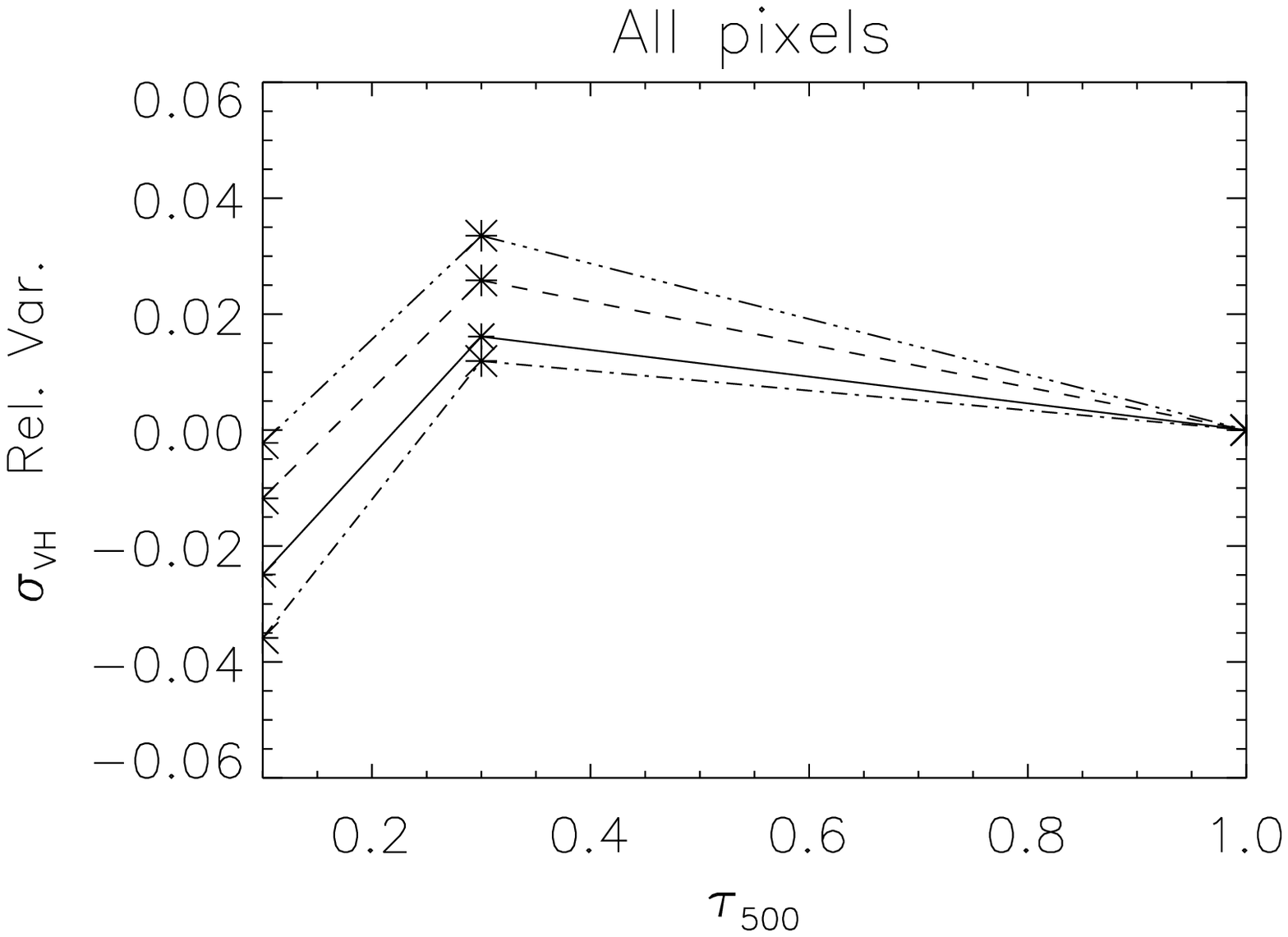}
\includegraphics[width=5cm]{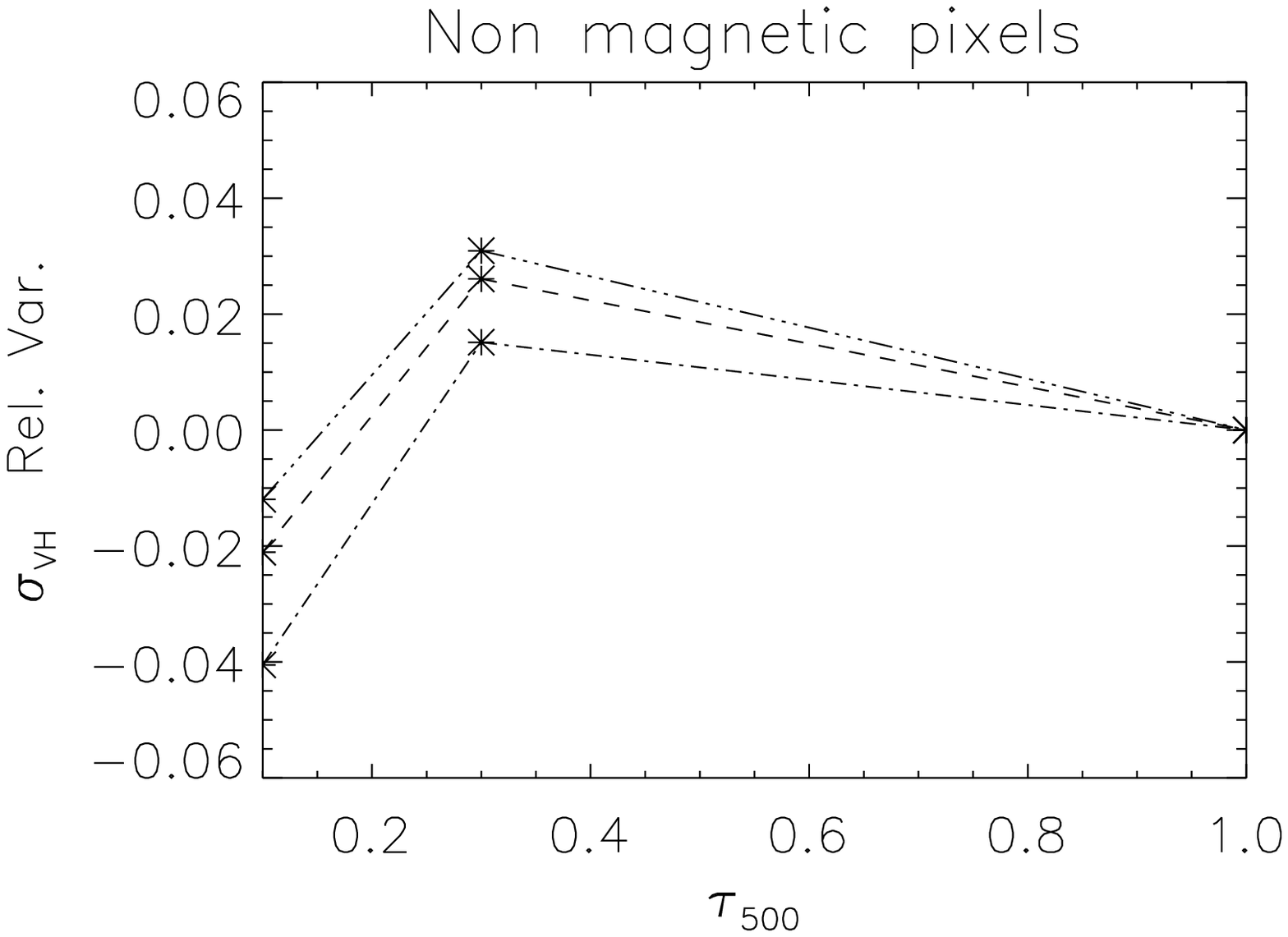}
\includegraphics[width=5cm]{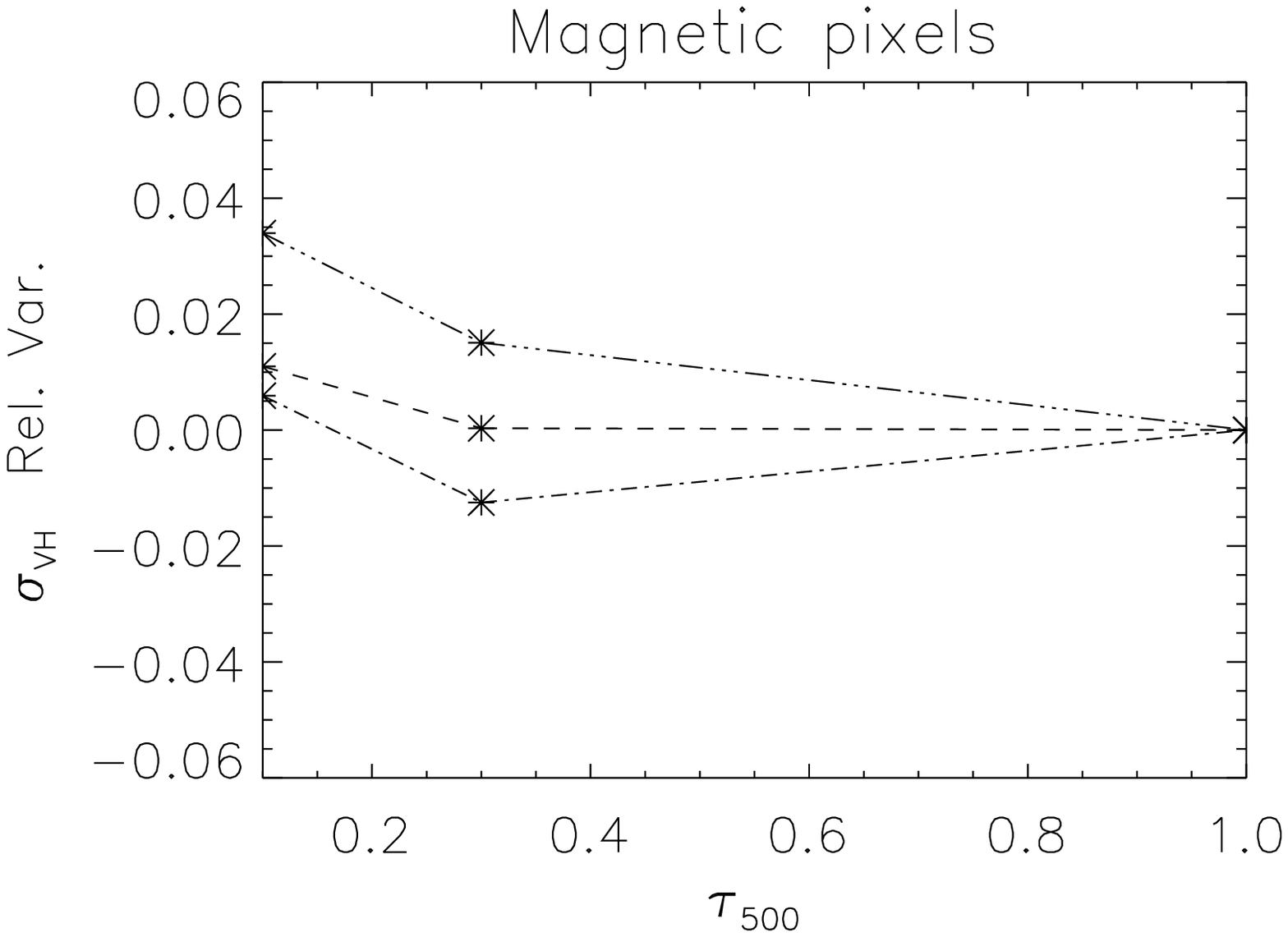}

\caption{Relative variation of the standard deviations of the velocity distributions with respect to standard deviation values at $\tau_{500}$ = 1. Top: Vertical velocities. Bottom: Horizontal velocities. Legend as in Fig. \ref{V_vert_tau1}.
\label{V_vert_rms_strati}  }

\end{figure}

In agreement with results obtained from HD simulations by other authors \cite[e.g.][]{asplund2000,steffen2007},  we found that the fluctuations of 
the vertical component of the velocity ($\sigma_{VV}$,hereafter) rapidly decreases with height. In particular, the plot in Fig.~\ref{V_vert_rms_strati} (top) shows that ($\sigma_{VV}$  of \textit{all pixels} and of \textit{non magnetic pixels} at $\tau_{500}=0.1$ are approximately 45$\%$ of the value at $\tau=1$ for all the environmental magnetic flux values investigated. In the case of \textit{magnetic pixels} the variations  are below 20$\%$ and, interestingly, are larger for larger environmental magnetic flux values. \textbf{These results suggest that the plasma velocities within magnetic structures overshoot to higher layers (velocity fluctuations decrease slower with height), although within magnetic structures located  in regions characterized by larger environmental magnetic flux velocity fluctuations decrease faster .  }

Instead, the standard deviations of the distributions of the horizontal velocities ($\sigma_{VH}$) do not show remarkable variations with height, as variations are within $5\%$ (bottom panel in Fig. \ref{V_vert_rms_strati}). Nevertheless, it is interesting to notice that for \textit{all pixels} and for the \textit{non magnetic pixels}, the values of the standard deviations do not show  monotonic variation, instead larger values are found at $\tau_{500}$ = 0.3.  This is a consequence of the adiabatic expansion of granules with height, and the consequent 'conversion' of velocities from vertical to horizontal. At higher layers the plasma has been decelerated and the edges of some granules show downflows instead of upflows.  The standard deviation of the distributions of the horizontal velocities of \textit{magnetic pixels}, instead, increases with height.  \citet{vogler2005a}, by the analysis of a 200 G MHD snapshot, found trends similar to the ones we found. These authors  ascribed the increase with height of the fluctuations of horizontal velocity of pixels characterized by higher values of magnetic field intensity,  to the correlation between horizontal velocities in the higher layers and downflow velocities surrounding magnetic structures in the lower layers of the photosphere. In all cases we find a small dependency of the variations of the standard deviations with height on the environmental magnetic flux, the variations being in general larger for higher environmental magnetic flux values.

\subsection{Temperature fluctuations}
\label{sec:temper}
\begin{figure*} 
\includegraphics[width=5cm,trim=1cm 1cm 1cm 1cm]{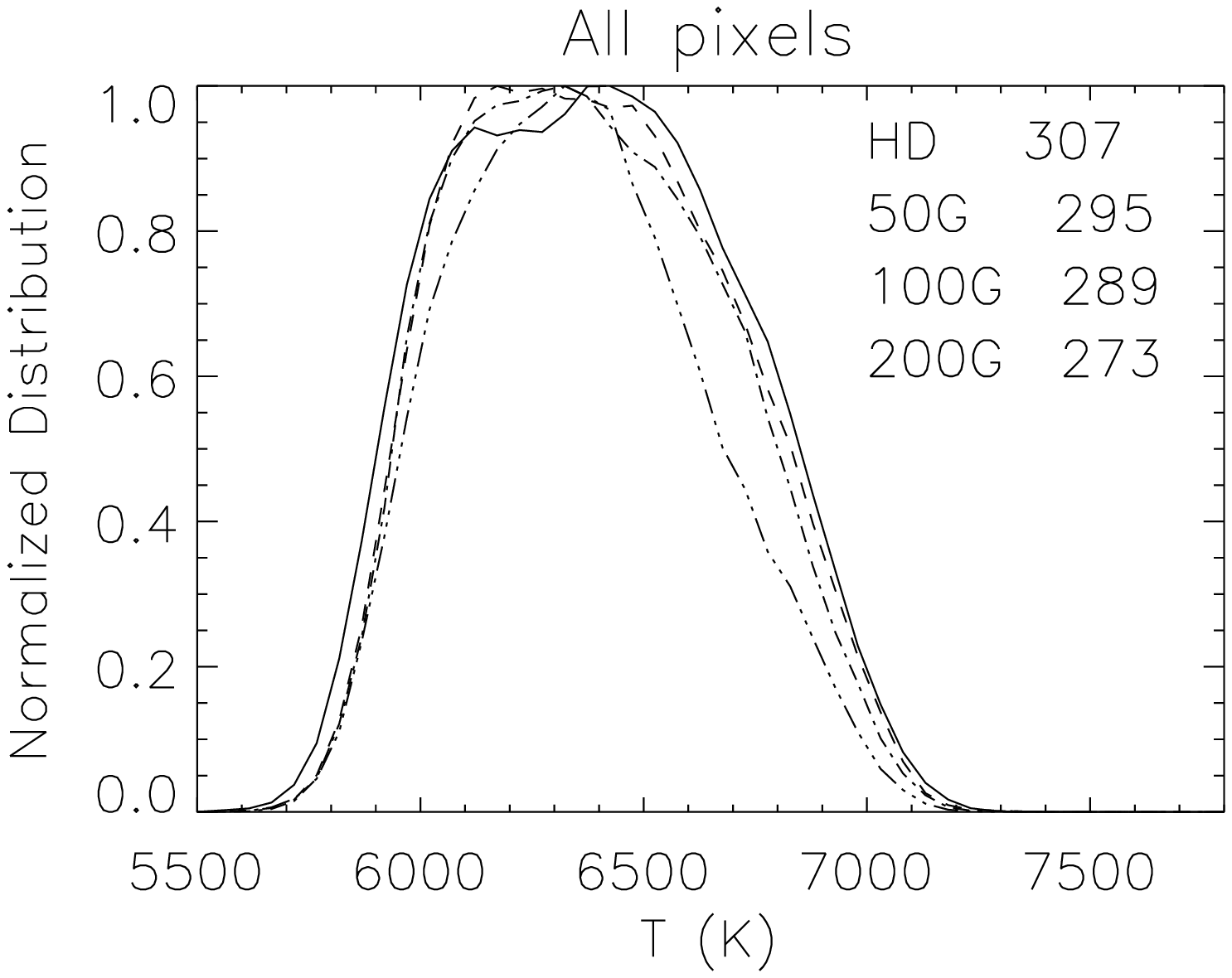}
\includegraphics[width=5cm,trim=1cm 1cm 1cm 1cm]{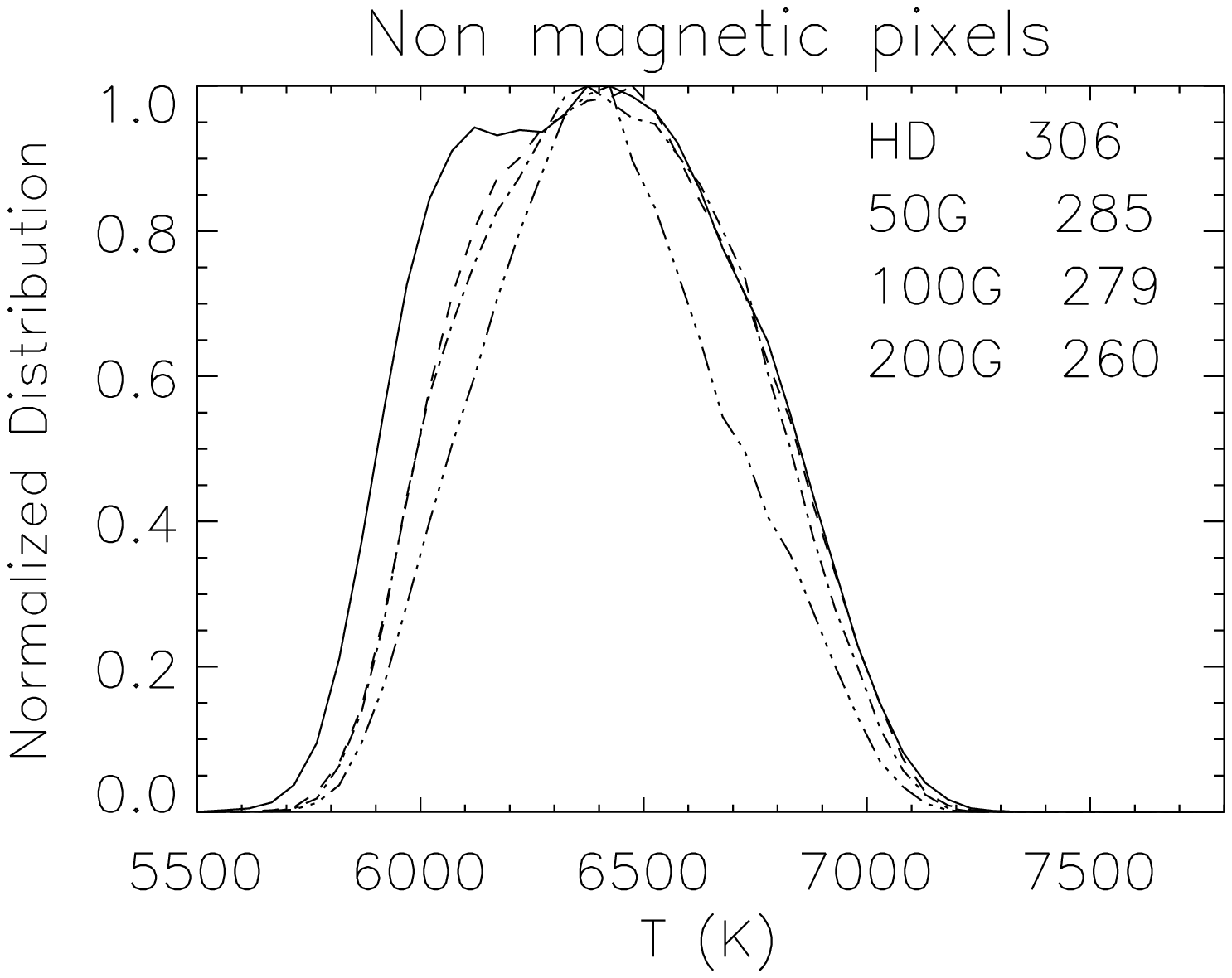}
\includegraphics[width=5cm,trim=1cm 1cm 1cm 1cm]{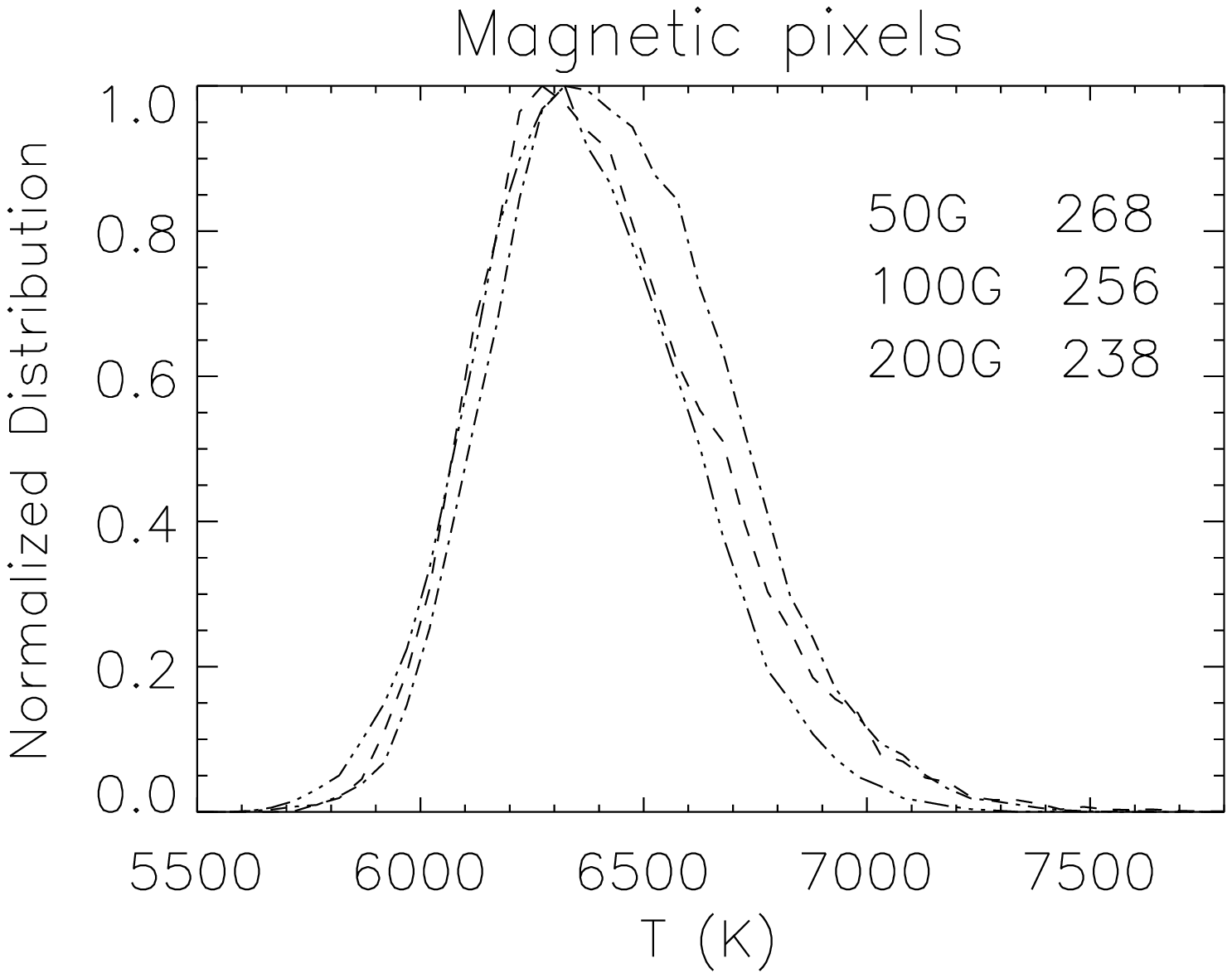}\\

\caption{Temperature distributions at $\tau_{500}$ = 1. Legend as in Fig.~ \ref{V_vert_tau1}.
\label{T_tau1}  }  
\end{figure*}

The distributions of temperature at $\tau_{500}$ = 1 and their standard deviations are shown in Fig. \ref{T_tau1}. We notice that some of the distributions are characterized by double peaks, which, as for the vertical velocities, are due to the presence of the  distinct populations of granules and intergranular lanes. Nevertheless, here the differences between the two peaks is less pronounced and indeed with the increase of the environmental magnetic flux the two peaks merge, so that in most of the MHD cases the distributions show a single broad peak instead of two. In all cases we observe a decrease of the amplitudes of the distributions with the increase of the environmental flux. As found for the velocities, standard deviations are smaller for \textit{magnetic pixels}.

\begin{figure}
\includegraphics[width=5cm]{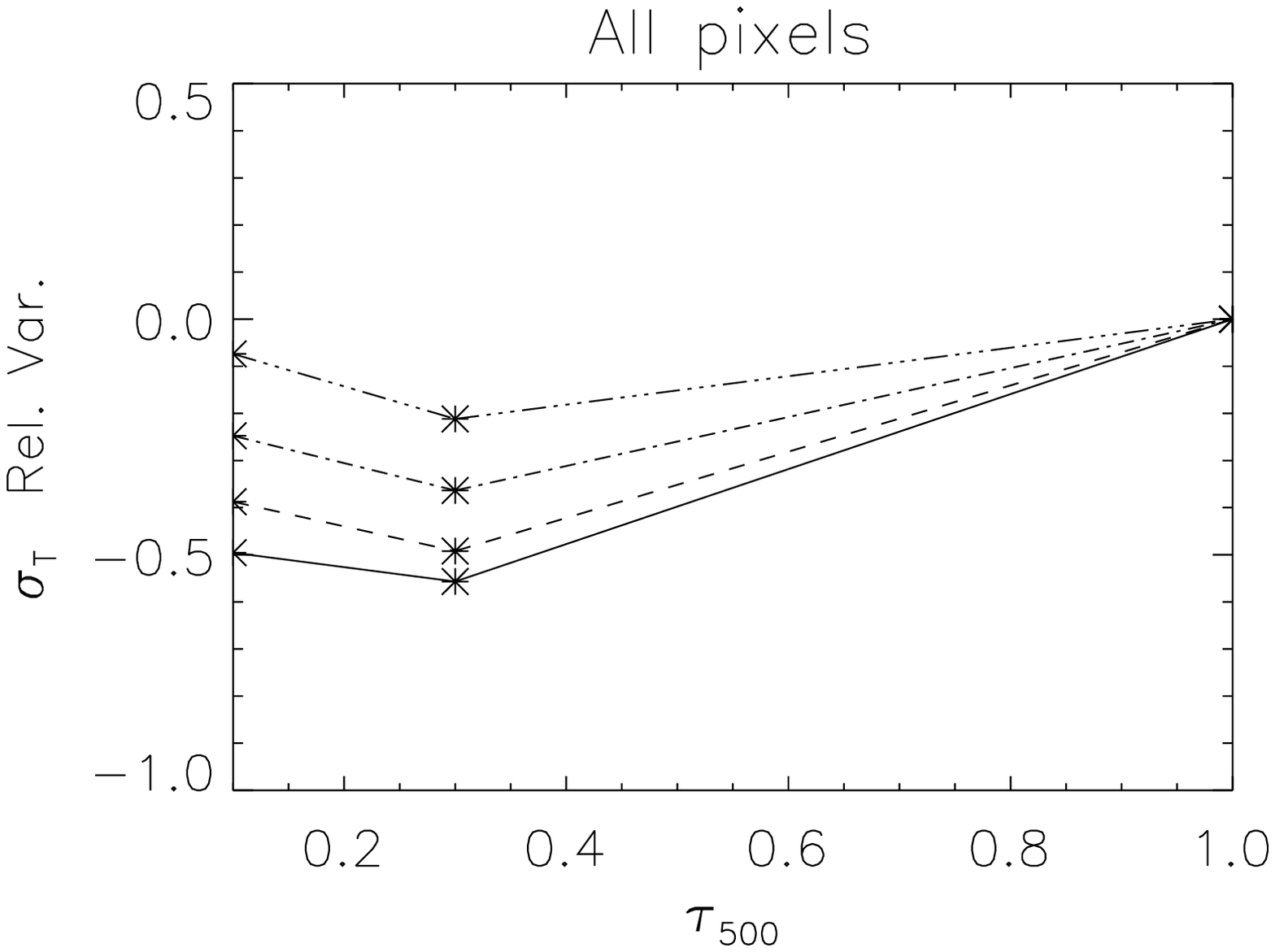}
\includegraphics[width=5cm]{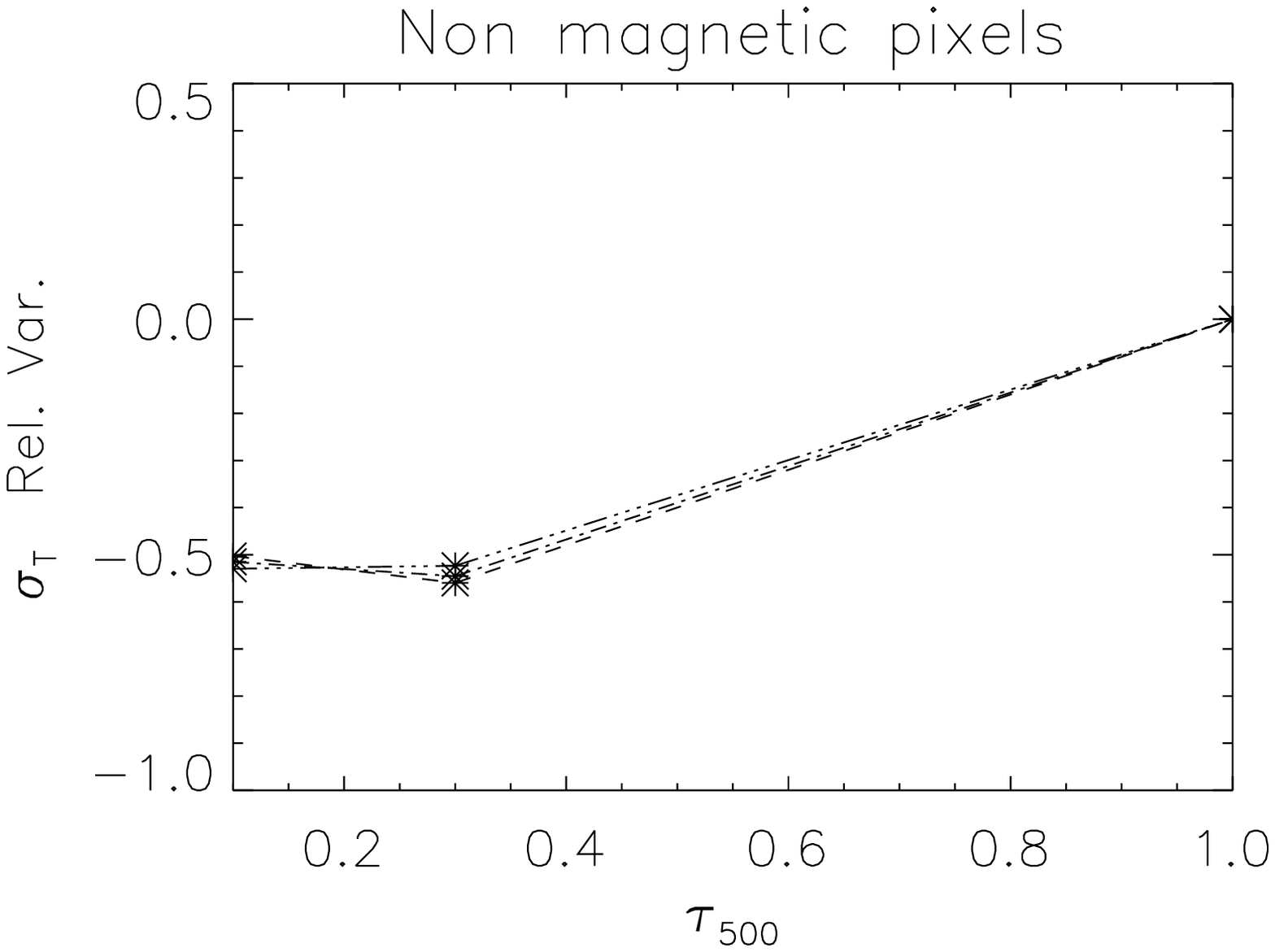}
\includegraphics[width=5cm]{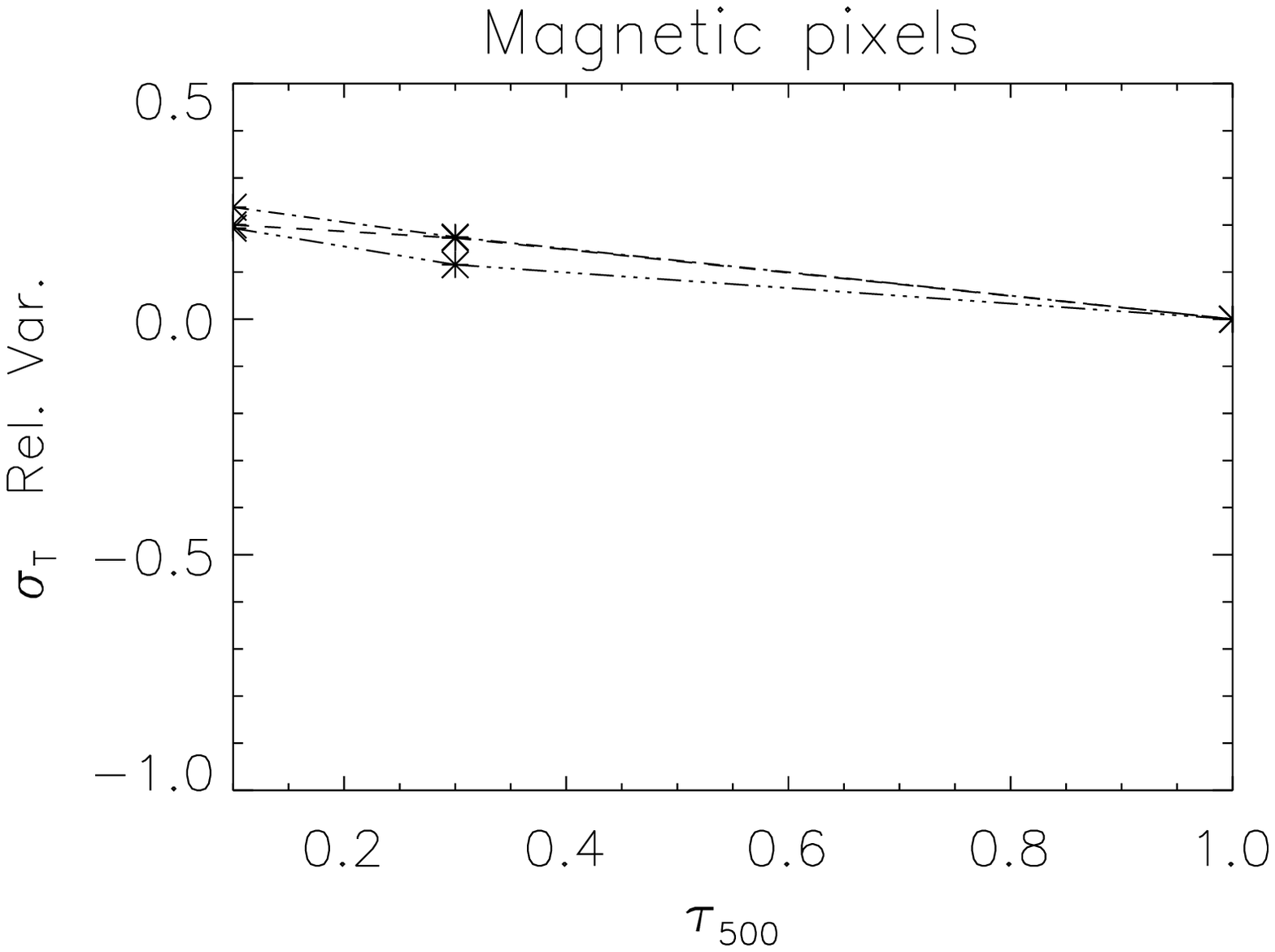}
\caption{Relative variation of standard deviation of temperature distributions. Legend as in Fig.~\ref{V_vert_tau1} 
\label{T_rms_strati}  }  

\end{figure}

 The relative variations of the standard deviations of temperatures ($\sigma_{T}$) with height are shown in Fig. \ref{T_rms_strati}.  When \textit{all pixels} are considered (black symbols), the variations are as high as $50\%$ in the case of HD snapshots; the amplitude of these variations is clearly reduced with the increase of the environmental magnetic flux. Actually, we notice that the trends are not monotonic and that instead a slight increase at $\tau_{500} = 0.1$ is observed. For \textit{non magnetic pixels}, instead, the decrease with height of the standard variation is monotonic and no dependence with the increase of the environmental magnetic flux is found. We conclude that the trend observed for \textit{all pixels} is due to the increase of the relative contribution of \textit{ magnetic pixels}  (red symbols), for which we observe in fact a monotonic increase with height. This increase is only slightly dependent on the amount of  the environmental magnetic flux.

\begin{figure} 
\includegraphics[width=7cm]{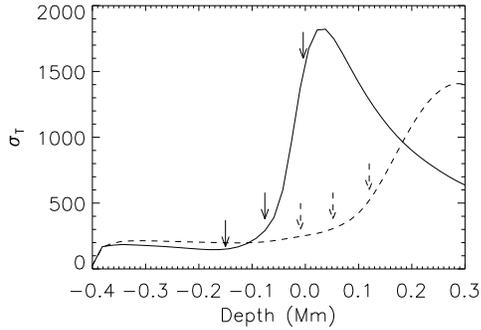}

\caption{Variation with geometric height of the standard deviation of temperature for \textit{non magnetic} (continuous line) and\textit{ magnetic pixels} (dashed line) in the case of the 200~G snapshots. The arrows indicate, from bottom to top, the heights at which the average optical depth at 500~nm is 1, 0.3 and 0.1, respectively.
\label{temp_fluct}  }  
\end{figure}

The decrease of the standard deviation of temperature with the decrease of optical depth in \textit{non magnetic pixels} is a consequence of the decrease of the vigor of convective motions with height. The increase of the temperature fluctuations for \textit{magnetic pixels} (which, as already noted in Sec.~\ref{sec:simu}, are not isolated pixels but instead appear within larger magnetic features), instead, is mostly a result of the reduction of the opacity within magnetic structures. In fact, at higher heights in the photosphere horizontal radiative effects become more and more important in determining the temperature, especially within magnetic  features (see Sec.~\ref{Sec:discuss}); as a consequence,  equal (vertical) optical depth surfaces do not trace anymore equal temperature surfaces (as happens in the deepest layers) and the temperature fluctuations increase with the decrease of the optical depth \citep[see also discussions in][] {uitenbroek2011}.  The plot in Fig. \ref{temp_fluct} shows indeed that the standard deviations of temperature decrease with geometric height in both \textit{magnetic} and \textit{non magnetic pixels}.

\subsection{Continuum intensity contrast}
\label{sec:contr}
\begin{figure}
\includegraphics[width=7cm]{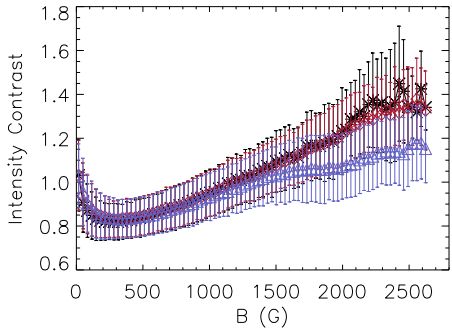}

\caption{Variation of the continuum intensity contrast at 608 nm with the increase of the absolute intensity of the vertical component of the magnetic field at $\tau_{500}$ = 1. Symbols denote results obtained from different values of environmental magnetic flux. Asterisks-Black: 50~G. Diamonds-Red: 100~G. Triangles-Blue: 200~G. 
\label{Contr_vs_B}  }

\end{figure}

We then investigated the relation between the magnetic field intensity and the  emergent continuum intensity at 608 nm for the various snapshots. We defined the continuum contrast as the ratio between the emergent intensity and the average of the emergent continuum intensities obtained from the HD snapshots. 
Figure \ref{Contr_vs_B} shows the variation of the contrast with the increase of the absolute intensity of the vertical component of the local magnetic field at $\tau_{500}$ = 1. For clarity, results have been averaged in 80 equally spaced bins; error bars denote the standard deviation values in each bin.

We notice that the contrast-magnetic field intensity relation is similar for all the MHD snapshots up to approximately 1100~G. At larger field intensities the contrast at equal local field strength obtained for the 200~G snapshots is systematically lower than that obtained for the other MHD snapshots, although differences are still within the standard deviations.

These differences must be partially ascribed to the increase of the number of large magnetic features \citep[whose contrast is known to be smaller from magnetic flux tube models, e.g.][]{spruit1976} with the increase of the environmental magnetic flux. To investigate the effects of the amount of magnetic flux in the surrounding environment, we therefore compared the intensity contrast of features characterized by similar magnetic field strength and area singled out in the different MHD snapshots. Results reported in Tab. \ref{tabellaree} show the average intensity contrast of features having average vertical magnetic field intensity at $\tau_{500}$ = 1 in the range 1000-1500~G. Results are binned by the equivalent radius, that is the radius of a circle having the same area of the magnetic feature. We note that for smaller features (radius between 90 and 200~km) the contrast does not show  a clear trend with the increase of the environmental magnetic flux. We ascribe this to the fact that, on average, these smallest features have associated magnetic field intensity of about 1200~G, a value for which plot shown in Fig. \ref{Contr_vs_B} show that the contrast is only slightly dependent on the environmental magnetic flux. Instead, larger features (radius larger than 200~km) have on average field intensity of approximately 1380~G, a value for which differences of contrast between pixels from snapshots characterized by different environmental magnetic flux are appreciable. Indeed, results in Tab. \ref{tabellaree} show for these features a clear decrease of the intensity contrast with the increase of the environmental magnetic flux. 

Finally, we investigated the continuum rms contrast of \textit{non magnetic pixels} and found 17.4$\% $, 16.3$\%$, 16$\%$, 14.7$\%$ for the HD, 50~G, 100~G and 200~G simulations, respectively. Similarly, \citet{uitenbroek2007} found from the analysis of MHD snapshots that the contrast in the G-band continuum of non magnetic features decreases with the increase of the environmental magnetic flux.

\section{Comparison with observations}
\label{sec:compar}
Results presented in the previous section are in good qualitative agreement with those from observations.
The decrease of the average vertical velocities and of their fluctuations in the photosphere with the increase of the environmental magnetic flux in both\textit{ non magnetic} and \textit{magnetic pixel}s is in agreement with findings by \citet{baudin1997,ishikawa2007,narayanscharmer2010,romano2012,kobel2012}.
Note that these authors do not report double-peaked distributions. The reason is, on one hand, the fact that the double peaks arise from small-scale structures (typically corresponding to down flows) whose size (few tens of km) is well below the detection limit of modern telescopes (about 100 km). On the other hand, inspection of the shapes of the velocity distributions obtained at different heights (not shown) reveals that, at typical formation heights of cores of photospheric lines usually employed to derive vertical velocities,  the distributions present a single peak.\textbf{ It is also worth to note that, due to spatial resolution effects, vertical velocity distributions obtained from observations are usually narrower than those reported in Sec. 3.1.}
 The shapes of the distributions of horizontal velocities are in  agreement with those reported by other authors both in the case of granular motions \citep[e.g.][]{baudin1997,attie2009} or in the case of G-band bright points motions  \citep[e.g.][]{keys2011}. Moreover, the decrease of the shift of the peak of the distributions and of their width with the increase of the environmental magnetic flux is consistent with observational results obtained by \citet{baudin1997}.\textbf{ Likewise the vertical velocity distributions, horizontal velocity distributions obtained from observations are narrower and peak at smaller values than those presented in this paper. In this case, differences must be ascribed to spatial resolution effects, temporal cadence \citep{attie2009} and methodology \citep{verma2013}. }

\textbf{The variation with height of the fluctuations of the vertical velocities (Fig. \ref{V_vert_rms_strati}) indicate that, in agreement with results presented by \citet{kostik2012}, within magnetic structures convection overshoots up to higher layers compared to non magnetic structures. We also found that within magnetic features located in higher magnetic environmental flux convection overshoots to lower layers with respect to magnetic features located in lower magnetic flux areas.  Apart from results reported by \citet{kostik2012},  to the best of our knowledge literature still lacks observational studies aimed at investigating whether the velocity reversal occurs at different heights in quiet and in active regions.}

A comparison of temperature fluctuations with results from observations is not straightforward, as this is not a direct measurable quantity. For observations in the lower layers of the atmosphere we can assume as a first approximation  that temperature and temperature fluctuations are correlated with the continuum intensity contrast \citep{uitenbroek2013}. Under this assumption, the reduction of the average temperature and of its fluctuations with the increase of the environmental magnetic flux is in good agreement with results reported by e.g. \citet{title1992,montagne1996,baudin1997,kobel2012,kostik2012} for granulation and by e.g. \citet{ishikawa2007,kobel2011,romano2012} for magnetic features. Indeed, we also found a decrease of the continuum intensity rms contrast of\textit{ non magnetic pixels} with the increase of the environmental magnetic flux.   

To the best of our knowledge, literature about comparison of variations of temperature (which can be performed through inversions of lines intensity profiles), or intensity contrast fluctuations with height of features embedded in different environments is still scarce. Measurements obtained from the analysis of intensity along spectral lines show that, in general, rms contrast of granulation shows a minimum around 100~km \citep[e.g.][]{espagnet1995} in the photosphere and then increases in the higher layers. \citet{baudin1997} found an increase of the rms contrast with height (they compared intensity contrast in the lower photosphere and at a height of approximately 140~km) for granulation located in quiet and active areas. Curiously, they found that the relative increase of the rms contrast is larger in the magnetically  active area. Our results show instead a rapid decrease of temperature fluctuations with optical depth, and a slight increase at $\tau_{500}$ = 0.1 for \textit{non magnetic pixels} in all MHD snapshots. Differences between observations and our results must be mostly ascribed to the broadening of the response functions of the line wings and cores usually employed for investigations of the rms contrasts with height in observations. We note in fact that, in agreement with our findings (see Fig. \ref{temp_fluct}), results from inversions of lines report a rapid decrease of the temperature fluctuations at heights below 200 km and then a slight increase at the higher layers \citep[e.g.][]{puschmann2005}.     

Finally, results shown in Fig.~\ref{Contr_vs_B} show that the continuum intensity contrast of pixels where magnetic field intensity is larger than approximately 1100~G depends on the amount of the magnetic flux of the region they are embedded in. Note that the intensity contrast-magnetic field relation we obtained for the 200 G snapshots is in good agreement with results reported by \citet{rohrbein2011} and \citet{danilovic2013} through the analysis of a 200 G MHD snapshot. Those authors showed that, once the spatial resolution of simulations is reduced to mimic realistically observational conditions, the dependence is similar to that reported in observations \citep[e.g.][]{kobel2011,schnerr2011,kobel2012}. We can therefore conclude that plots in Fig. \ref{Contr_vs_B} are in good agreement with observational results by \citet{kobel2011} and \citet{kobel2012} that show a different intensity contrast-magnetic field relation in quiet and active regions. 
Results in Tab. \ref{tabellaree} show more clearly that,\textbf{ in agreement with observations \citep[e.g.][]{lawrence1993,ishikawa2007,kobel2011,romano2012,feng2013},} photometric properties of magnetic features characterized by similar area and average magnetic field strength depend on the environment they are located in, their continuum intensity contrast being larger in quiet regions with respect to active ones.

\section{The importance of the temperature gradient}
\label{Sec:discuss}
In the previous section we have shown that the physical properties of convective and magnetic structures depend on the amount of magnetic flux of the environment in which they are embedded. These differences  must be mostly ascribed to the inhibition of convection and the reduction of the average opacity induced by the presence of the magnetic field.
In particular, the presence of  magnetic features which have associated kG field strength reduces both the average density and temperature, thus reducing the average opacity; moreover, it inhibits convection, thus causing a reduction of the plasma velocities. The two effects have opposite consequences on the temperature gradients: the reduction of opacity, which dominates at optical depths close to unity and smaller, reduces the temperature gradient; the inhibition of convection, which dominates at optical depths larger than one, steepens the temperature gradient \citep[c.f.r.][]{abbett1997}.

\begin{figure} 
\includegraphics[width=8cm,trim=1cm 1cm 1cm 0.4cm]{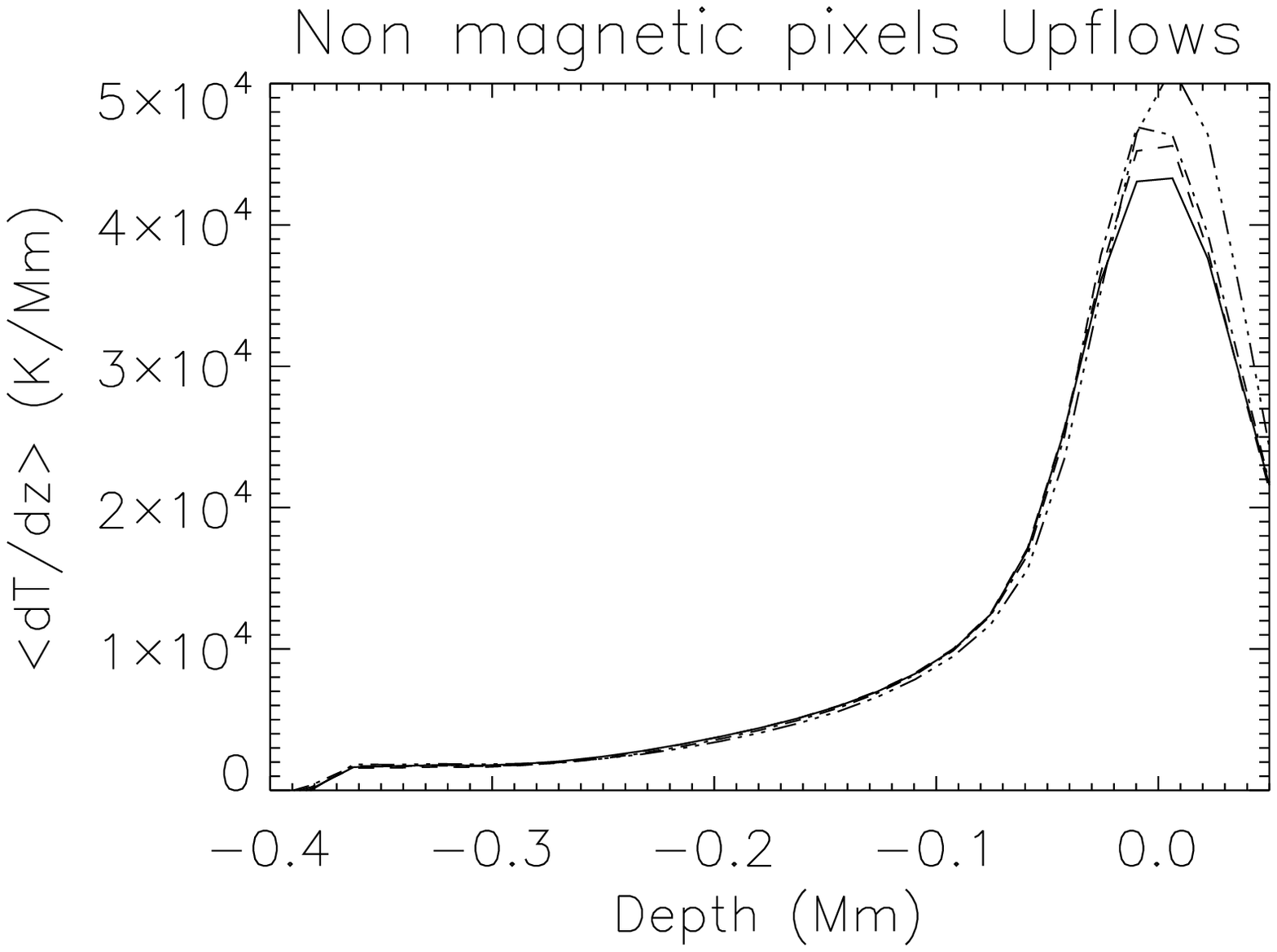}\\
\includegraphics[width=8cm,trim=1cm 1cm 1cm 0.4cm]{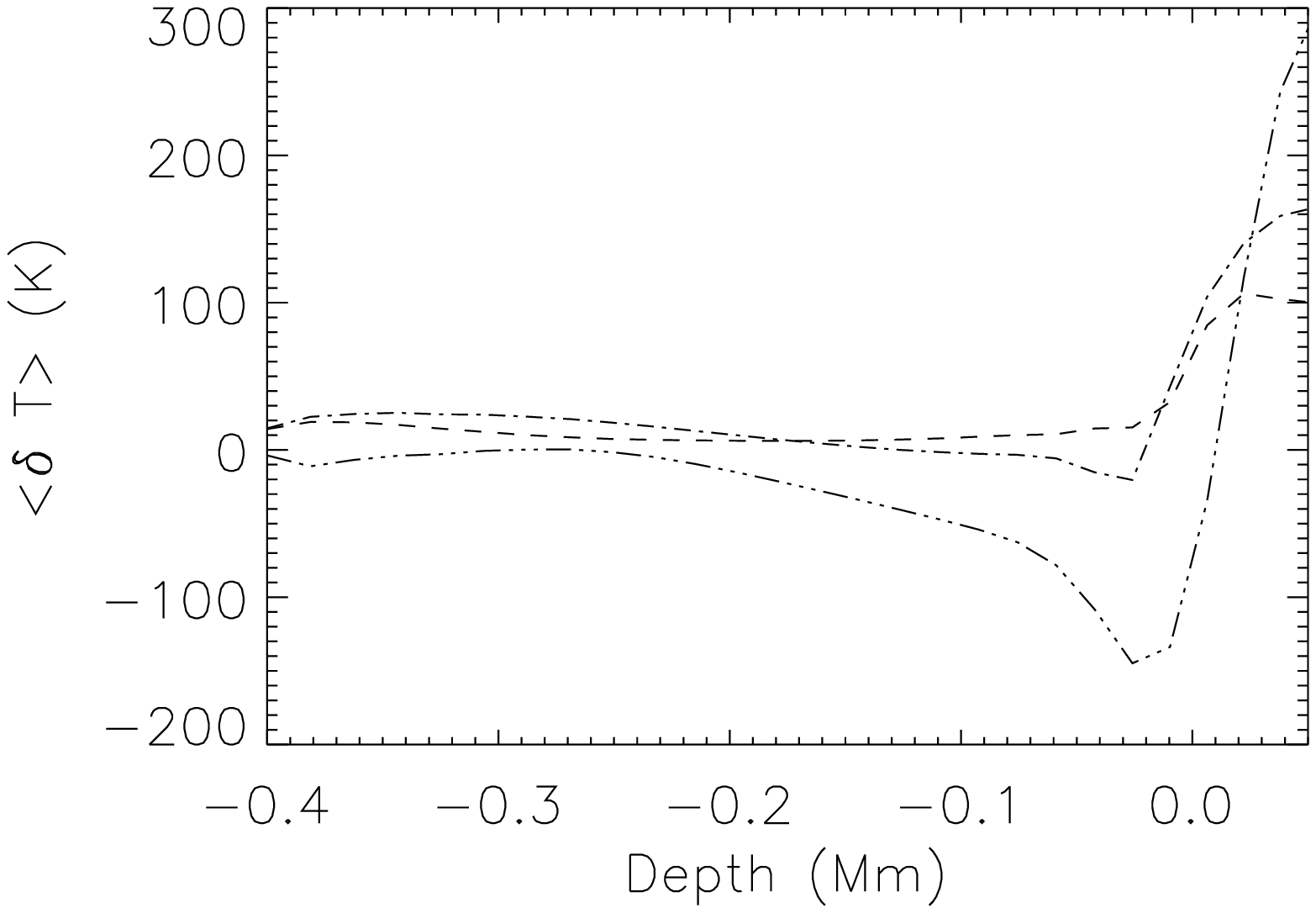}

\caption{Properties of average temperature stratification of \textit{non magnetic} pixels. Top: average temperature gradient. Bottom: difference between the average temperature stratification of the MHD simulations and the HD one. The zero position of the optical depth corresponds to the height of average optical depth unity in HD snapshots. Legend as in Fig.\ref{V_vert_tau1}  
\label{temperatures}  }  
\end{figure}

 This is illustrated in Fig. \ref{temperatures}, which shows the average temperature gradients (top) and the differences between the average temperature stratifications of the MHD snapshots and the HD ones (bottom) for \textit{non magnetic pixels}; to reduce the selection effects introduced by the fact that magnetic field concentrations are mostly located in downflow regions (see Sec. \ref{sec:conv}), the plots show the properties of the temperature stratification of upflow regions (that is of those pixels whose vertical velocity at $\tau_{500}$ = 1 is positive) only. The plots show clearly that in photospheric regions the temperature decreases with the increase of the environmental magnetic flux.

\begin{figure} 
\includegraphics[width=8cm,trim=1cm 1cm 1cm 0.4cm]{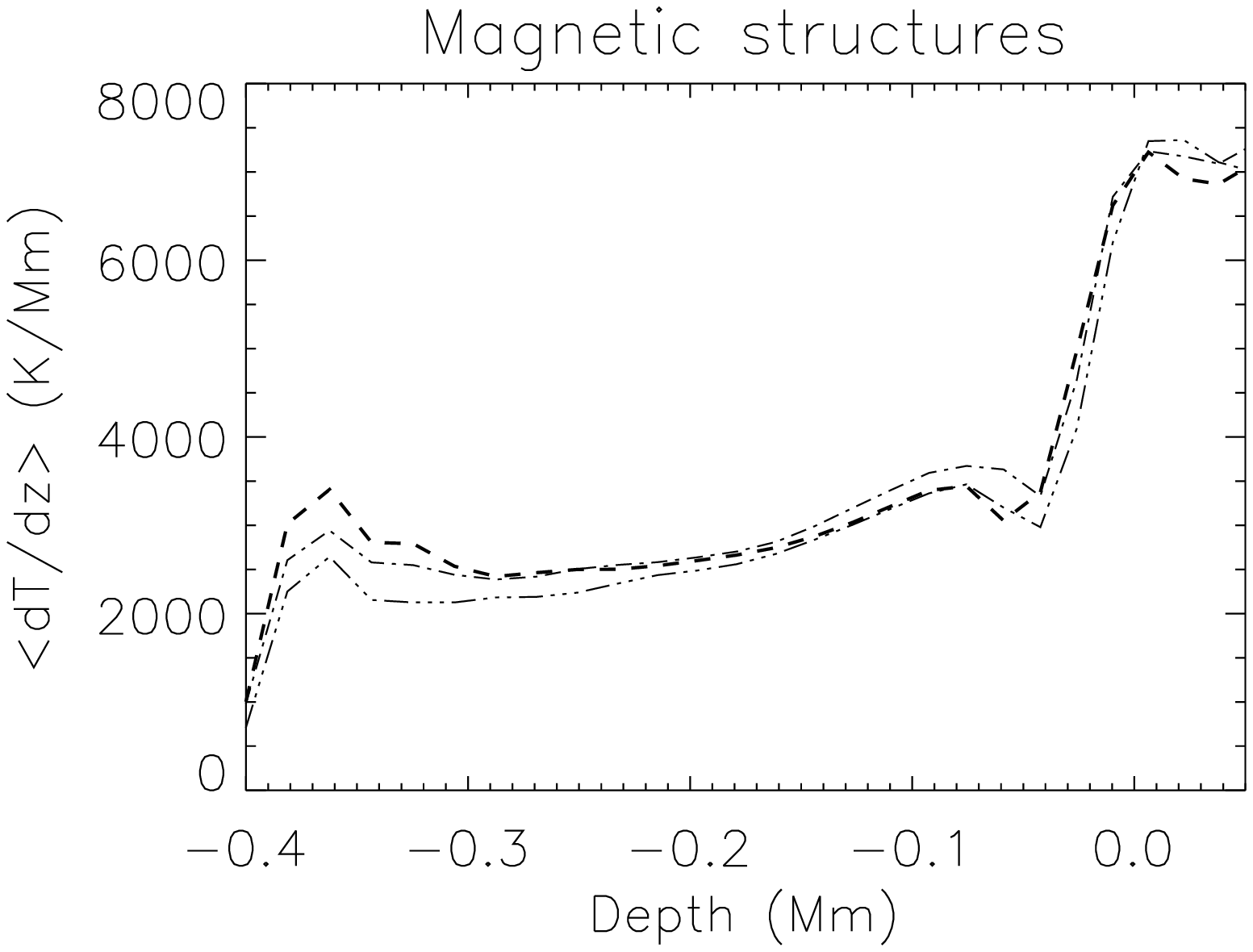}\\
\includegraphics[width=8cm,trim=1cm 1cm 1cm 0.4cm]{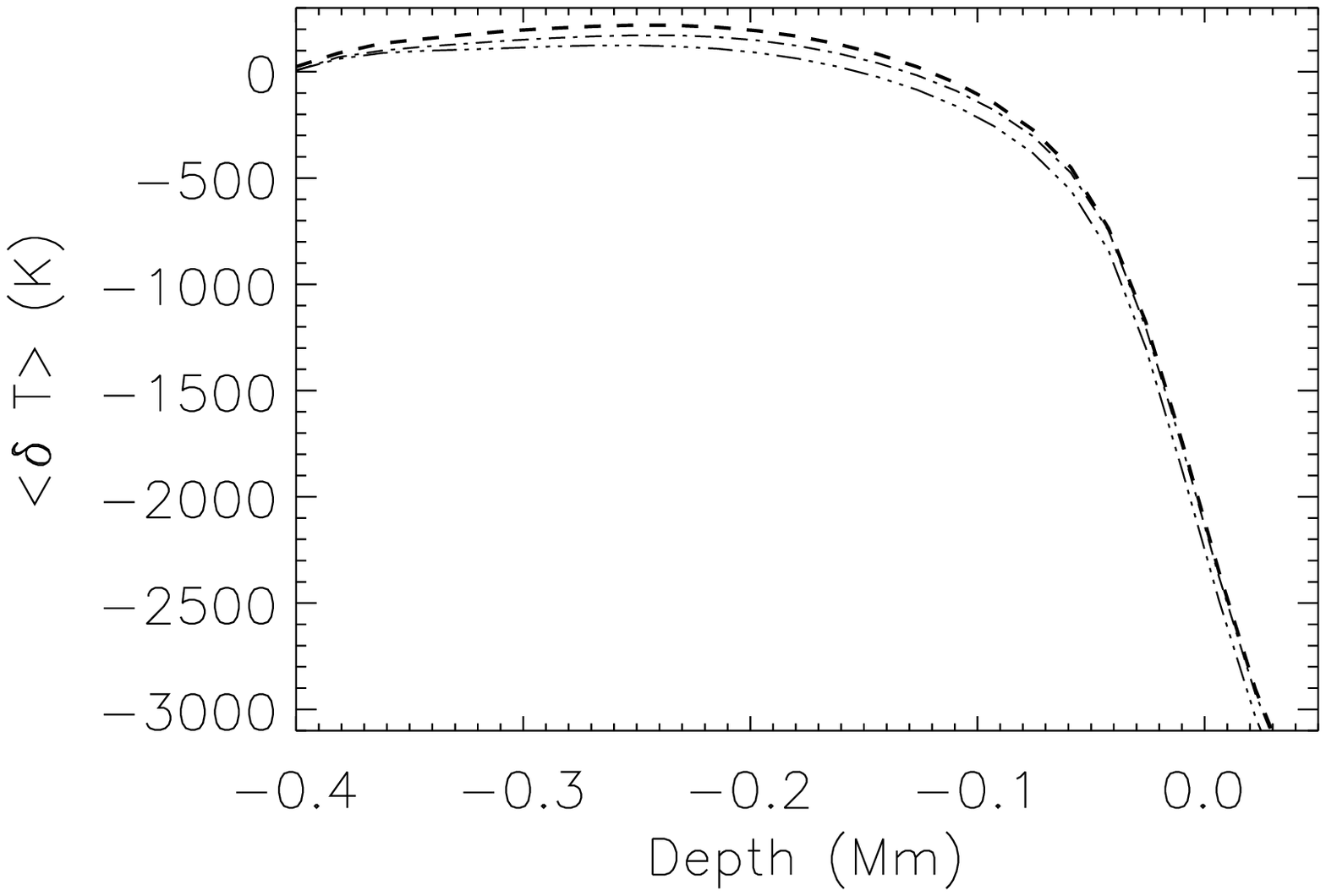}

\caption{Properties of average temperature stratification of magnetic structures having average magnetic field in the range 1000-1500 G and equivalent radius in the range 200-300~ km. Top: average temperature gradient. Bottom: difference between the average temperature stratifications of the magnetic features and of the HD snapshots. Legend as in Fig.\ref{V_vert_tau1}. For clarity results from the 50~G simulations are marked with a thicker line.  
\label{temperatures_mag}  }  
\end{figure}

The change of the temperature stratification of \textit{non magnetic pixels} also affects the stratification of\textit{ magnetic pixels}. In fact, flux tube models show that, at heights where the horizontal optical depth is comparable and larger than the size of the tube, the temperature stratification within a magnetic structure is determined, through radiative effects,  by the temperature of the surrounding plasma \citep[e.g.][]{criscuoli2009,steiner1998,pizzo1993}. Therefore, even the temperature of magnetic structures decreases with the increase of the environmental magnetic flux. 
This is confirmed by plots in Fig.\ref{temperatures_mag}, which show the same quantities as in Fig. \ref{temperatures} for magnetic features having field intensity between 1000 and 1500 G and equivalent radius between 200 and 300 km (see Sec.~\ref{sec:contr}). In particular, plot in bottom panel shows that in photospheric layers the difference of temperatures of the considered magnetic features embedded in different environment is at the most few hundred K, the temperature decreasing with the increase of the environmental magnetic flux. This explains the difference of photometric properties of magnetic features found in observations and in our simulations.
Plot in the top panel of Fig. \ref{temperatures_mag} shows that the temperature gradient of the selected magnetic features  slightly decreases with the increase of the environmental magnetic flux. Magnetic features embedded in regions characterized by lower magnetic flux are therefore characterized by higher temperatures and higher temperature gradients. As a consequence they experience larger radiative losses toward the upper layers, which favor larger downward motions \citep{rajaguru2000}. This explains the differences of vertical velocities observed within magnetic structures embedded in different environments, as well as the fact that magnetic features preserve their velocity structure  up to higher layers with respect to quiet regions. 

Finally, it is worth to note that a comparison of the plots reported in this section with those in \citet{fabbian2010, fabbian2012} and \citet{criscuoliuitenbroek2013}, which show the difference between the average temperature stratifications of the MHD and the HD snapshots versus the optical depth, indicate that the average temperature stratification in the photosphere ($\tau_{500} \leq$ 1) is mostly determined by the presence of magnetic structures.

\section{Summary and Conclusions}
We have analyzed and compared results from HD and MHD simulations characterized by different amount of average magnetic flux (environmental magnetic flux) to show that dynamic, thermal and photometric properties of both convective and magnetic structures depend on the environment in which they are embedded. 
We have shown that the dynamics of convection (vertical and horizontal motions) within and around magnetic features is more and more suppressed with the increase of the environmental magnetic flux. Both horizontal and vertical motions propagate differently toward the higher layers of the atmosphere  within non magnetic (regions where the vertical component of the magnetic field intensity at $\tau_{500}$ = 1 is lower than 50~G) and magnetic structures  (regions where the vertical component of the magnetic field intensity  at $\tau_{500}$ = 1 is higher than 800 G) in a manner that is slightly dependent on the amount of the environmental magnetic flux. In particular, within magnetic features vertical velocity fluctuations decrease more slowly with height than non magnetic features and their decrease is larger in regions characterized by higher environmental magnetic flux, \textbf{thus suggesting that convection overshoots  to higher layers within magnetic features.} Fluctuations of horizontal velocities instead increase monotonically toward higher layers of the atmosphere in the case of magnetic features, while they show a non monotonic trend in non magnetic features. Average temperatures and their fluctuations at $\tau_{500}$ = 1 of both non-magnetic and magnetic structures decrease with the increase of the environmental magnetic flux. Finally, we have showed that photometric continuum contrast of magnetic features characterized by similar average magnetic field intensity and similar area decreases with the increase of the environmental magnetic flux.

These results are in qualitative agreement with those from observations (see Sec.~\ref{sec:compar}). 

We have shown that the differences in photometric and dynamic properties found between features embedded in different environments  are due to the inhibition of convection induced by the presence of the magnetic field, which reduces the plasma velocities and modifies the temperature stratification of both granulation and magnetic features.

As already noted in Sec.~\ref{sec:intro}, these differences are still not taken into account in solar irradiance reconstruction models. For instance, we have found that even modest variations of the average magnetic flux can induce variations of the continuum intensity contrast of granulation. If identification methods employed in irradiance reconstructions  to single out classes of magnetic and quiet regions on solar disk images are not sensitive to such small variations of magnetic flux, then the reconstruction would "miss" a component. Similarly, the contribution of magnetic regions as faculae might not be properly modeled. In fact, irradiance reconstructions that model the radiative properties of magnetic features based on estimates of the magnetic  filling factor only \citep[e.g.][]{ball2011}, do not take into account that the temperature stratification, and therefore the radiative properties, of  magnetic elements are also determined by the properties of the surrounding plasma.  Models that employ intensity emission in chromospheric lines to identify magnetic features \citep[e.g.][]{fontenla2011},  suffer instead from the fact that different temperature stratifications can produce similar radiative emissions \citep{uitenbroek2011}, especially if differences between the temperature stratifications are larger in the lower photosphere and smaller above, as results reported in Sec.~\ref{Sec:discuss} indicate.
 All these aspects require further modeling, especially for what concerns the reconciliation of results from 3D MHD simulations and 1D static atmosphere models employed in irradiance reconstructions. In \citet{criscuoliuitenbroek2013} we employ this same set of (M)HD snapshots to investigate the effects of the changes of temperature gradient  induced by the increase of the environmental magnetic flux on spectral irradiance.   
  
Finally, it is worth to note that, since convection excites oscillatory motions along flux tubes, and that the propagation of waves within magnetic features depends  on the temperature stratification within the tubes \citep[e.g.][]{rajaguru2000,routh2010},  results presented in this paper could contribute to explain the differences of properties of oscillatory motions observed in network and faculae \citep[e.g.][and references there in]{chitta2012, khomenko2013}. Because the geometric properties of the magnetic field in the upper layers of the photosphere and chromosphere take an important role in determining the propagation of waves, 3D numerical simulations extending to layers higher than those explored in the present study are needed to investigate this issue.

\acknowledgements
 The author is grateful to H. Uitenbroek for reading the manuscript and discussing its content. The author also thanks Elena Khomenko for providing the series of snapshots of magnetoconvection simulations which were calculated using the computing resources of the MareNostrum (BSC/CNS, Spain)
and DEISA/HLRS (Germany) supercomputer installations. 


\begin{thebibliography}{}
\bibitem[Abbett et al.(1997)]{abbett1997}
Abbett, W.P., Beaver, M., Davids, B., Georgobiani, D., Rathbun, P., Stein, R. F.  1997, \aap, 480, 395


\bibitem[Asplund et al.(2000)]{asplund2000}
Asplund, M., Ludwig, H.G., Nordlund, \AA., Stein, R. F. 2000, \aap, 359, 669

\bibitem[Attie et al.(2009)]{attie2009}
Attie, R., Innes, D.E., Potts, H.E. 2009, \aap, 493, 13

\bibitem[Ball et al.(2011)]{ball2011}
	Ball, W.T., Unruh, Y.C., Krivova, N.A., Solanki, S., Harder, J.W. 2011, \aap, 530, 71

\bibitem[Baudin et al.(1997)]{baudin1997}
Baudin, F., Molowny-Horas, R., Koutchmy, S.  1997, \aap, 326, 842

\bibitem[Brandenburg et al.(2005)]{brandenburg2005}
Brandenburg, A., Chan, K. L., Nordlund, \AA., Stein, R. F. 2005, AN, 326, 681

\bibitem[Cattaneo (2003)]{cattaneo2003}
Cattaneo, F., Emonet, T., Weisss, N. 2003, \apj{}, 588, 1183

\bibitem[Chitta et al.(2012)]{chitta2012}
Chitta, L., Jain, R., Kariyappa, R., Jefferies, S.M. 2012, \apj, 744, 98

\bibitem[Criscuoli \& Rast(2009)]{criscuoli2009}
Criscuoli, S., Rast, M. P. 2009, \aap, 495, 691


\bibitem[Criscuoli et al.(2013)]{criscuoli2013}
Criscuoli, S., Ermolli, I., Uitenbroek, H., Giorgi, F. 2013, \apj, 763, 144

\bibitem[Criscuoli \& Uitenbroek(2013)]{criscuoliuitenbroek2013}
Criscuoli, S., Uitenbroek, H. 2013 in preparation

\bibitem[Danilovic et al.(2013)]{danilovic2013}
Danilovic, S., R\"{o}hrbein, D., Cameron, R.H., Sch\"{u}ssler, M. 2013 \aap, 550, 118


\bibitem[Ermolli et al.(2013)]{ermolli2013}
Ermolli, I., Matthes, K., Dudok de Wit, T., et al. 2013, ACP, 13.3945E

\bibitem[Espagnet et al.(1995)]{espagnet1995}
Espagnet, O., Muller, R., Roudier, T., Mein, N., Mein, P.  1995, \aap S, 109, 79

\bibitem[Fabbian et al.(2010)]{fabbian2010}
Fabbian, D., Khomenko, E., Moreno-Insertis, F., Nordlund, \AA.  2010, \apj, 724, 1536

\bibitem[Fabbian et al.(2012)]{fabbian2012}
Fabbian, D., Moreno-Insertis, F., Khomenko, E.,  Nordlund, \AA.  2012, \apj, 548, A35


\bibitem[Fontenla et al.(2011)]{fontenla2011}
Fontenla, J., Harder, J., Livingston, W., Snow, M., Woods, T. 2011, \jgr, 11620108F


\bibitem[Fr\"{o}hlich(2011)]{frohlich2011}
Fr\"{o}hlich, C. 2011, SSRv, 133F

\bibitem[Harder et al.(2009)]{harder2009}
Harder, J.W., Fontenla, J. M., Pilewskie, P., Richard, E.C., Woods, T.N. 2009, Geophys. Res. Lett., 36, L07801

\bibitem[Ishikawa et al.(2007)]{ishikawa2007}
Ishikawa, R., Tsuneta, S., Kitakoshi, Y., et al. 2007, \aap, 472, 911


\bibitem[Lawrence et al.(1993)]{lawrence1993}
Lawrence, J.K., Topka, K.P., Jones, H.P. 1993, JGR, 9818911L


\bibitem[Leka \& Steiner (2001)]{lekasteiner2001}
Leka, K.D., Steiner, O. 1998, \apj, 552, 354


\bibitem[Keller \& Koutchmy(1991)]{keller1991}
Keller, C.U., Koutchmy, S. 1991, \apj, 379, 751

\bibitem[Keys et al.(2011)]{keys2011}
Keys, P. H., Mathioudakis, M., Jess, D. B. et al. 2011, \apj, 740L, 40

\bibitem[Khomenko \& Calvo Santamaria(2013)]{khomenko2013}
Khomenko, E., Calvo Santamaria, I. 2013, JPhCS, 440a2048K 

\bibitem[Kobel et al.(2011)]{kobel2011}
Kobel, P., Solanki, S. K., Borrero, J. M. 2011, \aap, 531, A112

\bibitem[Kobel et al.(2012)]{kobel2012}
Kobel, P., Solanki, S. K., Borrero, J. M. 2012, \aap, 542, A96

\bibitem[Kostik \& Khomenko(2012)]{kostik2012}
Kostik, R., Khomenko, E. 2012, \aap, 545, A22

\bibitem[Morinaga et al.(2008)]{morinaga2008}
Morinaga, S., Sakurai, T., Ichimoto, K., et al. 2008, \aap, 481L, 29

\bibitem[Montagne et al.(1996)]{montagne1996}
Montagne, M., M\"{u}ller, R., Vigneau, J. 1996, \aap, 311, 304

\bibitem[Narayan \& Scharmer(2010)]{narayanscharmer2010}
Narayan, G., Scharmer, G. 2010, \aap, 524, A3

\bibitem[Nesis et al.(2006)]{nesis2006}
Nesis, A., Hammer, R., Roth, M., Schleicher, H. 2006, \aap, 451, 1081

\bibitem[Nordlund \& Galsgaard(1995)]{nordlund1995}
Nordlund, \AA ., Garlsgaard, K. 1995,   
Tech. Rep., Astron. Observ., Copenhagen
Univ., http://www.astro.ku.dk/ aake/papers/95.eps.gz

\bibitem[Pizzo et al.(1993)]{pizzo1993}
Pizzo, V. J., MacGregor, K. B., Kunasz, P. B. 1993 \apj, 404, 788

\bibitem[Puschmann et al.(2005)]{puschmann2005}
Puschmann, K.G., Ruiz Cobo, B., V\'{a}zquez, M., Bonet, J. A., Hanslmeier, A. 2005, \aap, 441, 1157

\bibitem[Rajaguru \& Hasan(2000)]{rajaguru2000}
Rajaguru, S.P., Hasan, S.S. 2000, \aap, 544, 522

\bibitem[Routh et al.(2010)]{routh2010}
Routh, S., Musielak, Z. E., Hammer, R. 2010, \apj, 709, 1297


\bibitem[R\"{o}hrbein et al.(2011)]{rohrbein2011}
R\"{o}hrbein, D. Cameron, R., Sch\"{u}ssler, M. 2011 \aap, 532, 140

\bibitem[Romano et al.(2012)]{romano2012}
Romano, P., Berrilli, F., Criscuoli, et al. 2012, \solphys, doi: 10.1007/s11207-012-9942-7


\bibitem[Schnerr \& Spruit(2011)]{schnerr2011}
Schnerr, R.S., Spruit, H.C. 2011, \aap, 532, A136

\bibitem[Feng et al.(2013)]{feng2013}
Feng, S., Deng, L., Yang, Y., Ji, K. 2013, Ap\&SS, in press

\bibitem[Spruit(1976)]{spruit1976}
Spruit, H.C. 1976, \solphys, 50, 269

\bibitem[Steffen (2007)]{steffen2007}
Steffen, M.  2007, IAUS, 239, 36

\bibitem[Stix (1989)]{stix}
Stix,  M.  1989, The Sun: An Introduction (Berlin and Heidelberg: Springer).


\bibitem[Steiner et al.(1998)]{steiner1998}
Steiner, O., Grossman-Doerth, U., Kn\"{o}lker, M., Sch\"{u}ssler, M. 1998, \apj, 495, 468

\bibitem[Title et al.(1992)]{title1992}
Title, A.M., Topka, K. P., Tarbell, T.D., et al. 1992, \apj, 393, 782

\bibitem[Uitenbroek(2002)]{uitenbroek2002}
Uitenbroek, H. 2002, \apj, 565, 1312

\bibitem[Uitenbroek(2003)]{uitenbroek2003}
Uitenbroek, H. 2003, \apj, 592, 1225

\bibitem[Uitenbroek(2007)]{uitenbroek2007}
Uitenbroek, H., Tritschler, A., Rimmele, T. 2007, \apj, 668, 586

\bibitem[Uitenbroek \& Criscuoli(2011)]{uitenbroek2011}
Uitenbroek, H., Criscuoli, S. 2011, \apj, 736, 69

\bibitem[Uitenbroek \& Criscuoli(2013)]{uitenbroek2013}
Uitenbroek, H., Criscuoli, S. submitted to \apj

\bibitem[Verma et al.(2013)]{verma2013}
Verma, M., Steffen, M., Denker, C.  2013 \aap, 555, 136


\bibitem[V\"{o}gler et al.(2005a)]{vogler2005a}
V\"{o}gler, A., Shelyag, S., Sch\"{u}ssler, M., et al. 2005a, \aap,429, 335

\bibitem[V\"{o}gler et al.(2005b)]{vogler2005b}
V\"{o}gler, A.  2005b, MmSAI, 76, 842


\end{thebibliography}

\clearpage


\begin{table}
\begin{center}
\begin{tabular} {cccc}
\tableline
MHD & \multicolumn{3}{c} {Equivalent Radius}\\
\tableline
\tableline
 & 90-200 km  & 200-300 km & 300-400 km \\
\tableline

50G  &   1.03$\pm$0.07 	  & 1.10$\pm$0.06  & 1.08$\pm$0.03 \\
100G & 	1.05$\pm$0.07    & 1.07$\pm$0.07  & 1.06$\pm$0.04 \\
200G &   1.03$\pm$0.075   & 1.04$\pm$0.06  & 1.03$\pm$0.07
\end{tabular}
\end{center}
\caption{\label{tabellaree} Intensity Contrast of features whose average absolute intensity of vertical component of the magnetic field at $\tau_{500}$ = 1 is  between 1000 and 1500~G.}
\end{table}

\end{document}